\documentclass[twocolumn]{aastex62}
\turnoffedit
\usepackage[all]{nowidow}
\usepackage{stix}
\usepackage{xspace}
\usepackage{hyperref}
\usepackage{todonotes}
\usepackage{graphicx}
\usepackage{enumitem}

\newcommand{\mjup}{\ensuremath{M_\mathrm{Jup}}\xspace}
\newcommand{\teff}{\ensuremath{T_{\mathrm{eff}}}\xspace}
\newcommand{\logg}{\ensuremath{\log(g)}\xspace}
\newcommand{\fluxunit}{\ensuremath{\mathrm{ergs}\,\mathrm{cm^{-2}}\,\mathrm{s^{-1}}\,\mathrm{\micron^{{-1}}}}}
\graphicspath{{./}{figures/}}
\shorttitle{HD106906b Time-Resolved Observations}
\shortauthors{Zhou et al.}

\begin{document}

\title{Cloud Atlas: High-precision HST/WFC3/IR Time-Resolved Observations of Directly-Imaged Exoplanet HD106906b}

\correspondingauthor{Yifan Zhou}
\email{yifan.zhou@utexas.edu}

\author{Yifan Zhou}
\altaffiliation{Harlan J. Smith McDonald Observatory Fellow}
\affiliation{Department of Astronomy/Steward Observatory, The University of Arizona, 933 N. Cherry Avenue, Tucson, AZ, 85721, USA}
\affiliation{Department of Astronomy/McDonald Observatory, The University of Texas, 2515 Speedway, Austin, TX, 78712, USA}

\author{D\'aniel Apai}
\affiliation{Department of Astronomy/Steward Observatory, The University of Arizona, 933 N. Cherry Avenue, Tucson, AZ, 85721, USA}
\affiliation{Department of Planetary Science/Lunar and Planetary Laboratory, The University of Arizona, 1640 E. University Boulevard, Tucson, AZ, 85718, USA}
\affiliation{Earths in Other Solar Systems Team, NASA Nexus for Exoplanet System Science.}

\author{Luigi R. Bedin}
\affiliation{INAF – Osservatorio Astronomico di Padova, Vicolo dell'Osservatorio 5, I-35122 Padova, Italy}

\author{Ben W. P. Lew}
\affiliation{Department of Planetary Science/Lunar and Planetary Laboratory, The University of Arizona, 1640 E. University Boulevard, Tucson, AZ, 85718, USA}

\author{Glenn Schneider}
\affiliation{Department of Astronomy/Steward Observatory, The University of Arizona, 933 N. Cherry Avenue, Tucson, AZ, 85721, USA}

\author{Adam J. Burgasser}
\affiliation{Center for Astrophysics and Space Science, University of California San Diego, La Jolla, CA 92093, USA}

\author{Elena Manjavacas}
\affiliation{W. M. Keck Observatory, 65-1120 Mamalahoa Hwy. Kamuela, HI, 96743, USA}

\author{Theodora Karalidi}
\affiliation{Department of Physics, University of Central Florida, 4111 Libra Dr, Orlando, FL 32816}

\author{Stanimir Metchev}
\affiliation{Department of Physics \& Astronomy and Centre for Planetary Science and Exploration, The University of Western Ontario, London, Ontario N6A 3K7, Canada}
\affiliation{Department of Astrophysics, American Museum of Natural History, Central Park West at 79th Street, New York, NY 10024-5192, USA}

\author{Paulo A. Miles-P\'aez}
\altaffiliation{ESO Fellow}
\affiliation{European Southern Observatory, Karl-Schwarzschild-Stra{\ss}e 2, 85748 Garching, Germany}

\author{Nicolas B. Cowan}
\affiliation{Department of Earth \& Planetary Sciences and Department of Physics, McGill University, 3550 Rue University, Montr\'eal, Quebec H3A 0E8, Canada}

\author{Patrick J. Lowrance}
\affiliation{IPAC-Spitzer, MC 314-6, California Institute of Technology, Pasadena, CA 91125, USA}

\author{Jacqueline Radigan} \affiliation{Utah Valley University, 800 West University Parkway, Orem, UT 84058, USA}

\begin{abstract}
 HD106906b is an $\sim11\mjup$, $\sim 15$\,Myr old directly-imaged exoplanet orbiting at an extremely large distance from its host star. The wide separation (7.11\arcsec) between HD106906b and its host star greatly reduces the difficulty in direct-imaging observations, making it one of the most favorable directly-imaged exoplanets for detailed characterization. In this paper, we present HST/WFC3/IR time-resolved observations of HD106906b in the F127M, F139M, and F153M bands. We have achieved $\sim1\%$ precision in the lightcurves in all three bands. The F127M lightcurve demonstrates marginally-detectable (\edit1{$2.7\sigma$ significance}) variability with a best-fitting period of 4 hr, while the lightcurves in the other two bands are consistent with flat lines. We construct primary-subtracted deep images and use these images to exclude additional companions to HD106906 that are more massive than 4\mjup{} and locate at projected distances of more than $\sim500$ au. We measure the astrometry of HD106906b in two HST/WFC3 epochs and achieve precisions better than 2.5 mas. The position angle and separation measurements do not deviate from those in the 2004 HST/ACS/HRC images for more than $1\sigma$ uncertainty. We provide the HST/WFC3 astrometric results for 25 background stars that can be used as reference sources in future precision astrometry studies. Our observations also provide the first 1.4-\micron{} water band photometric measurement for HD106906b. HD106906b's spectral energy distribution and the best-fitting BT-Settl model have an inconsistency in the 1.4-\micron{} water absorption band, which highlights the challenges in modeling atmospheres of young planetary-mass objects.
\end{abstract}

\keywords{Planetary Systems --- planets and satellites: atmospheres --- methods: observational}

\section{Introduction}

Condensates clouds are central components of the atmospheres of brown dwarfs and exoplanets \citep[e.g.,][]{Ackerman2001,Morley2012,Marley2015}. Cloud opacity strongly impacts near-infrared (NIR) colors and spectra of these objects. Therefore, understanding cloud properties is critical to determining  fundamental properties and atmospheric compositions of substellar objects through emission and transmission spectroscopic observations \citep[e.g.,][]{Ingraham2014, Kreidberg2014a, Stevenson2016, DeWit2016, Samland2017}. Because brown dwarfs are available for direct spectroscopy and their observations are generally less challenging than spectroscopic observations of transiting exoplanets, cloud properties for brown dwarfs are more tightly constrained than for exoplanets through time-averaged spectroscopic \citep[e.g.,][]{Burgasser2008,Cushing2008,Stephens2009} and time-resolved \citep[e.g.,][]{Buenzli2012,Apai2013,Yang2016,Biller2017,Schlawin2017,Apai2017} observations. Directly-imaged exoplanets and planetary-mass companions \citep[e.g.,][]{Chauvin2004,Marois2008a,Marois2010,Macintosh2015a}, which overlap with transiting planets in mass and are suitable for high-quality time-series observations, are excellent targets for connecting condensate cloud studies of brown dwarfs and exoplanets.

HD106906b is an $11\pm2$ \mjup{}  mass exoplanet orbiting an F5V spectral-type star \citep{Bailey2013}. Based on spectroscopic analysis \citep{Bailey2013,Daemgen2017}, the planet has an effective temperature (\teff) of approximately 1,800~K and a spectral type of L2.5-3. The HD106906 system, at a distance of $103.3\pm0.4$ pc \citep{Gaia2016,Gaia2018}, is a member of the Lower Centaurus Crux association (99.8\% membership probability based on BANYAN-$\Sigma$, \citealt{Gagne2018} ), which itself is part of the Sco-Cen OB association. Based on its cluster membership, the age of the system is  $15\pm3$ Myr \citep[][]{Pecaut2016}. The planet has a wide separation of $7\arcsec.11\pm0\arcsec.03$ from its host star \citep{Bailey2013}, corresponding to a projected distance of $734\pm4$ au. Because of the planet's large angular separation from its host star, the incident flux from the bright host star does not contaminate that from the companion significantly, despite the large brightness contrast ($\Delta J=10.3\,\mathrm{mag}$). Therefore, HD106906b is among the most favorable exoplanets for atmospheric characterization \citep[e.g., ][]{Bailey2013,Kalas2015,Wu2016,Daemgen2017}.

Multi-wavelength photometric \citep{Bailey2013,Kalas2015,Wu2016} and spectroscopic \citep{Bailey2013,Daemgen2017} observations have been used to characterize HD106906b's atmosphere. In these observations, similar to many other young L-type planetary-mass objects (2M1207b, \citealt{Chauvin2004}, HR8799bcde, \citealt{Marois2008,Marois2010}, PSO J318, \citealt{Liu2013}), HD106906b appears reddened in its NIR colors compared to those of the field brown dwarfs of the same spectral type. The reddened NIR color is often associated with dusty atmospheres and thick condensate clouds \citep[e.g.,][]{Skemer2011,Barman2011,Bowler2013,Liu2016}. Time-resolved observations of these reddened objects have often found  them to be variable \citep[e.g.,][]{Biller2015,Zhou2016,Lew2016,Vos2017,Biller2017,Manjavacas2017,Zhou2019}. Several of these objects also demonstrate wavelength-dependent variability \citep[e.g.,][]{Zhou2016,Lew2016,Biller2017,Zhou2019}, of which the amplitude is higher in the shorter wavelength (e.g. $J$ band) than in the longer wavelength (e.g., $H$ band).  The most likely cause of the variability and its wavelength-dependency is heterogeneous clouds rotationally modulating the disk-integrated flux from the photosphere. Consequently, multi-wavelength NIR rotational modulation has become an effective tool to study condensate clouds, particular vertical cloud profiles and dust grain properties for brown dwarfs and planetary-mass objects \citep[e.g.,][]{Apai2013, Biller2017,Manjavacas2017,Schlawin2017,Manjavacas2019,Paez2019,Zhou2019}. High-precision time-resolved NIR observations can thus be an effective method to explore the cloud properties of HD106906b.

HD106906b's extremely wide orbit and its deviation from the host star's circumstellar disk plane pose challenges in explaining its formation \citep{Bailey2013,Kalas2015,Lagrange2016,Wu2016}. Disk fragmentation has difficulty forming a planet/companion with a mass as small as that of HD106906b \citep[e.g.,][]{Kratter2010}. High-contrast direct-imaging surveys strongly support core accretion as the formation pathway of planetary-mass companions with orbits smaller than 100~au \citep{Wagner2019,Nielsen2019}.  At a projected distance of more than 700~au from its host star \citep{Bailey2013}, it is unlikely for HD106906b to accrete enough material through \emph{in situ} core accretion. A $\sim21^{\circ}$ projected angle between the planet's position angle and the plane of its host star's disk \citep{Kalas2015} further argues against \emph{in situ} core accretion but suggests dynamical orbit evolution of this planet \citep[e.g.,][]{Marleau2019}. \edit1{The host HD106906 is likely to be a spectroscopic binary \citep{Lagrange2016,Wu2016,Rodet2017,DeRosa2019}, corroborating the scenario where the current planetary orbit is a consequence of dynamical interactions between the host and the planet.} \citet{DeRosa2019} discovered a close, near-coplanar stellar encounter with the HD106906 system, further supporting a conjecture of intense dynamical activity in the system's evolution history. Considering these evidence that suggests past dynamical evolution, it should not be surprising if HD106906b has an eccentric orbit. Therefore, astrometric constraints on the orbit of HD106906b will be critical for understanding the formation and evolution history of HD106906b.

In this paper, we analyze and discuss \emph{Hubble Space Telescope} Wide Field Camera 3 near-infrared channel (HST/WFC3/IR) observations of HD106906b in time-resolved direct-imaging mode. We present lightcurves of HD106906b in three bands that cover the 1.4~\micron{} water band and its the continuum. We look for variability in the lightcurves and use them to discuss the atmospheric and cloud properties of HD106906b. We also compare the relative astrometry of HD106906 system in the two WFC3 observations and in the HST Advanced Camera for Survey/High-Resolution Channel (ACS/HRC) observations, which were taken in 2004. The WFC3 and ACS/HRC observations together form a high astrometric precision  image series with 14 years baseline,  which can place tight constraints on the motion of HD106906b relative to its host star.

\section{Observations}

\begin{figure*}
  \centering
\plottwo{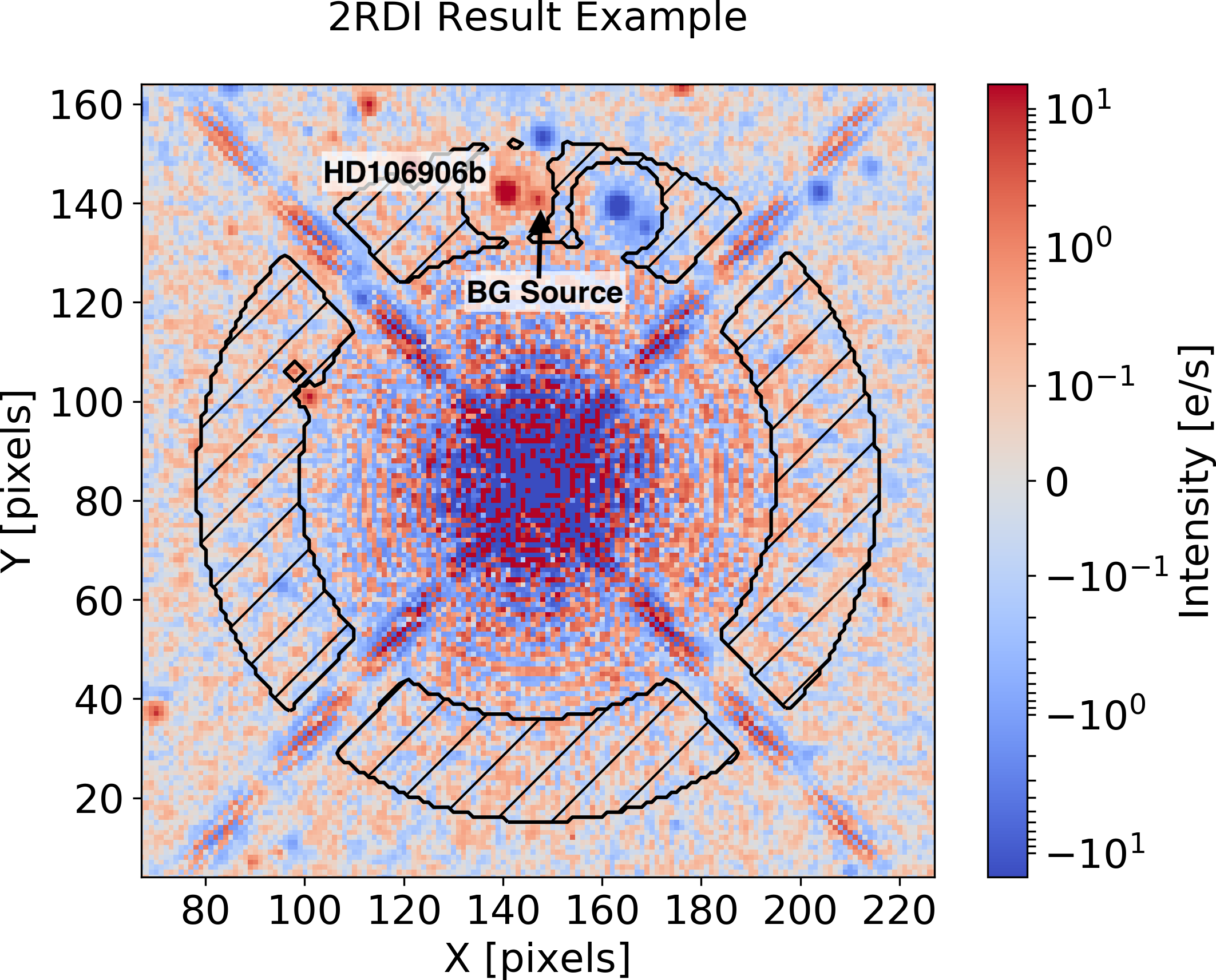}{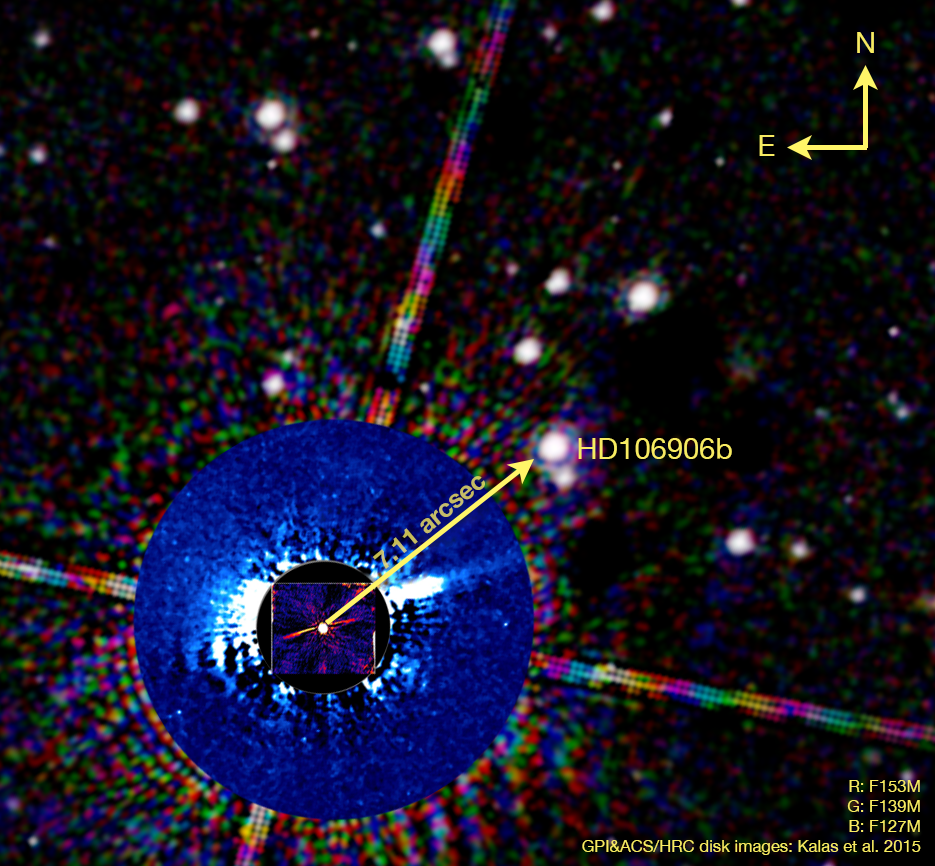}  
\caption{Direct-imaging observations of the HD106906 system. \emph{Left:} A demonstration of the two-roll differential imaging results. Red color represents signals from the original images and blue colored pixels are structures from the subtraction model images. Regions that are marked by hatches are used for optimizing the subtraction. \edit1{HD106906b and a nearby (in projection) uncataloged source (later identified as BG12) are marked in the figure. To avoid the uncataloged source contaminating the photometry for HD106906b, PSF fitting is carried out simultaneously for these two sources.} \emph{Right: An R (F153M) G (F139M) B (F127M) color composite image of HD106906.} Overlaid on the HST RGB composite are the false-color Gemini Planet Imager (inner most) and ACS/HRC (outer annulus) scattered light images \citep{Kalas2015} of the circumstellar disk. The circumstellar disk is not visible in the WFC3/IR images.}
  \label{fig:2rdi}
\end{figure*}

The HST/WFC3/IR observations of HD106906 \edit1{(MAST DOI: \href{http://archive.stsci.edu/doi/resolve/resolve.html?doi=10.17909/t9-13te-fp08}{10.17909/t9-13te-fp08})} are part of the HST Large Treasury program \emph{Cloud Atlas} (Program ID: 14241, PI: D. Apai). We observed HD106906 from 2016-01-29 20:45 to 2016-01-29 23:02 UTC in two consecutive HST orbits as part of the program's variability amplitude assessment survey (VAAS). We then used the same instrument set-up to re-visit the target from 2018-06-07 02:14 to 2018-06-07 12:35 UTC  in seven consecutive HST orbits as part of the deep look observations (DLO). Dithering was not applied during the observation to reduce systematics caused by flat field errors. The target was observed in F127M ($\lambda_{\mathrm{pivot}}=1.274\micron$, $\mathrm{FWHM}=0.07\micron$), F139M ($\lambda_{\mathrm{pivot}}=1.384\micron$, $\mathrm{FWHM}=0.07\micron$) and F153M ($\lambda_{\mathrm{pivot}}=1.533\micron$, $\mathrm{FWHM}=0.07\micron$) filters.  The filter selection allowed comparison of the modulations  in (F139M) and out  (F127M, F153M) of the 1.4 \micron{} water absorption band.  Exposure times were 66.4 seconds for the F127M and F153M observations and 88.4 seconds for the F139M observations. We alternated these three filters in every two or three exposures, and thus the lightcurves in the three filters are almost contemporaneous. 

The observations were designed to enable two-roll angular differential imaging for primary point spread function (PSF) star subtraction.  This technique was successfuly applied in HST high-contrast observations \citep[e.g.,][]{Zhou2016,Zhou2019,Paez2019}. Successive orbits alternately differed in celestial orientation angle 31 degrees apart, with odd (1, 3, 5, and 7), and even (2, 4, 6) numbered orbits respectively at the same orientations. Subtracting images taken in the odd orbits from those  taken in the even orbits (or vice versa) removes the primary star PSF (in the absence of systemmatics to the level of the photon noise) but conserves the companion PSF (Figure \ref{fig:2rdi}).  

HD106906 was also observed by HST/ACS/HRC on 2004-12-01 UTC (PID: 10330, PI: H. Ford). The 2004 ACS/HRC observations include two identical 1,250 seconds direct-imaging exposures in the ACS F606W  band. We use results  from these observations \citep{Bailey2013,Kalas2015} to extend the temporal baseline for our astrometric analysis.

\section{Data Reduction}

\subsection{Time-Resolved Photometry}
We start our time-resolved photometry with the \texttt{flt} files produced by the CALWFC3 pipeline. Our photometric data reduction has four steps: data preparation, primary star subtraction, PSF-fitting photometry, and lightcurve systematics removal. Data reductions for lightcurves in the three filters are independent. Therefore, the four reduction steps are applied to observations in the three filters in parallel. 

In the data preparation step, we organize the bad-pixel-masked and sky-subtracted images into data cubes. First, we make bad pixel masks and remove the sky background. Pixels that have data quality flags 4 (bad detector pixel), 16 (hot pixel), 32 (unstable response), and 256 (full-well saturation) are identified as ``bad pixels'', masked out, and excluded from subsequent analyses. We then further examine images by eye to identify and mask out remaining spurious pixels. There are no bad pixels within a 5-pixel radius aperture centered on HD106906b and thus the effect of bad pixels on the photometry is negligible. To remove the sky background, we first draw circular masks around all visible point sources in the field of view and then apply a five-iteration  sigma-clip (threshold: $5\sigma$) to exclude remaining bright pixels. We take the median value of the unmasked pixels as sky background and subtract it from every image. The background-subtracted images and the associated bad pixel masks are sorted in chronological order and stored in data cubes.

We then apply two-roll differential imaging \citep[2RDI, e.g.,][]{Lowrance1999,Song2006} to subtract the PSF of the primary star.  Images taken with the first telescope roll are subtraction template candidates for images taken with the second telescope roll and vice versa. We measure the primary star positional offset in each image using two-dimensional cross-correlation and align the primary star PSFs with bi-linear interpolation shift. We refine image registration by least $\chi^{2}$ optimization in the diffraction spike regions that are caused by the secondary mirror support structures. We then select the best subtraction template from all available candidate images.  Each subtraction template candidate is linearly scaled to minimize the squared summed subtraction residuals in the original$-$template image in an annulus around HD106906A (Figure \ref{fig:2rdi}). The best subtraction template is the one that results in the least subtraction residuals. Finally, we subtract the best templates from the original images to obtain primary subtracted images (Figure \ref{fig:2rdi}). 

HD106906b's flux intensity is measured by PSF fitting to the primary-subtracted images. Details of the PSF-fitting procedures can be found in \citet{Zhou2019}. We construct $9\times$ over-sampled PSFs using the TinyTim PSF modeling software \citep{Krist1995}. Free parameters for the model PSFs are the centroid coordinates, HST secondary mirror displacement, and the amplitude of the PSF. We optimize these parameters using a maximum likelihood method combined with Markov Chain Monte Carlo (MCMC) algorithms \citep[MCMC performed by \texttt{emcee},][]{Foreman-Mackey2012}. Aperture correction for each filter band is done through PSF fitting photometry as we normalize the model PSF to flux within an infinitely large aperture. We note that there is an uncataloged source (discussed later in \S\ref{sec:othersources}) that is only 0.87\arcsec away from HD106906b. To avoid this source contaminating the photometry of HD106906b, we also create PSF models for it and conduct PSF fitting for HD106906b and this uncataloged source simultaneously. 

Finally, we correct the lightcurve systematics and estimate the photometric uncertainty.  For WFC3/IR lightcurves, charge trapping related ramp effect is the major component of lightcurve systematic noise. We use RECTE \citep{Zhou2017} to model and remove the ramp effect systematics from the lightcurves. Our implementation of the ramp effect removal procedure follows \citet{Zhou2019}, in which details of the application of RECTE in time-resolved direct imaging observations are provided. We calculate ramp profiles by feeding the entire time series into RECTE and forward-modeling the charge trapping systematics. The model ramp profiles are divided from the lightcurves to correct the systematics.  We estimate the photometric uncertainty by combining  photon noise, detector readout noise, and dark current in pixels that are used for the measurements. 

\begin{figure}[!h]
  \centering
  \plotone{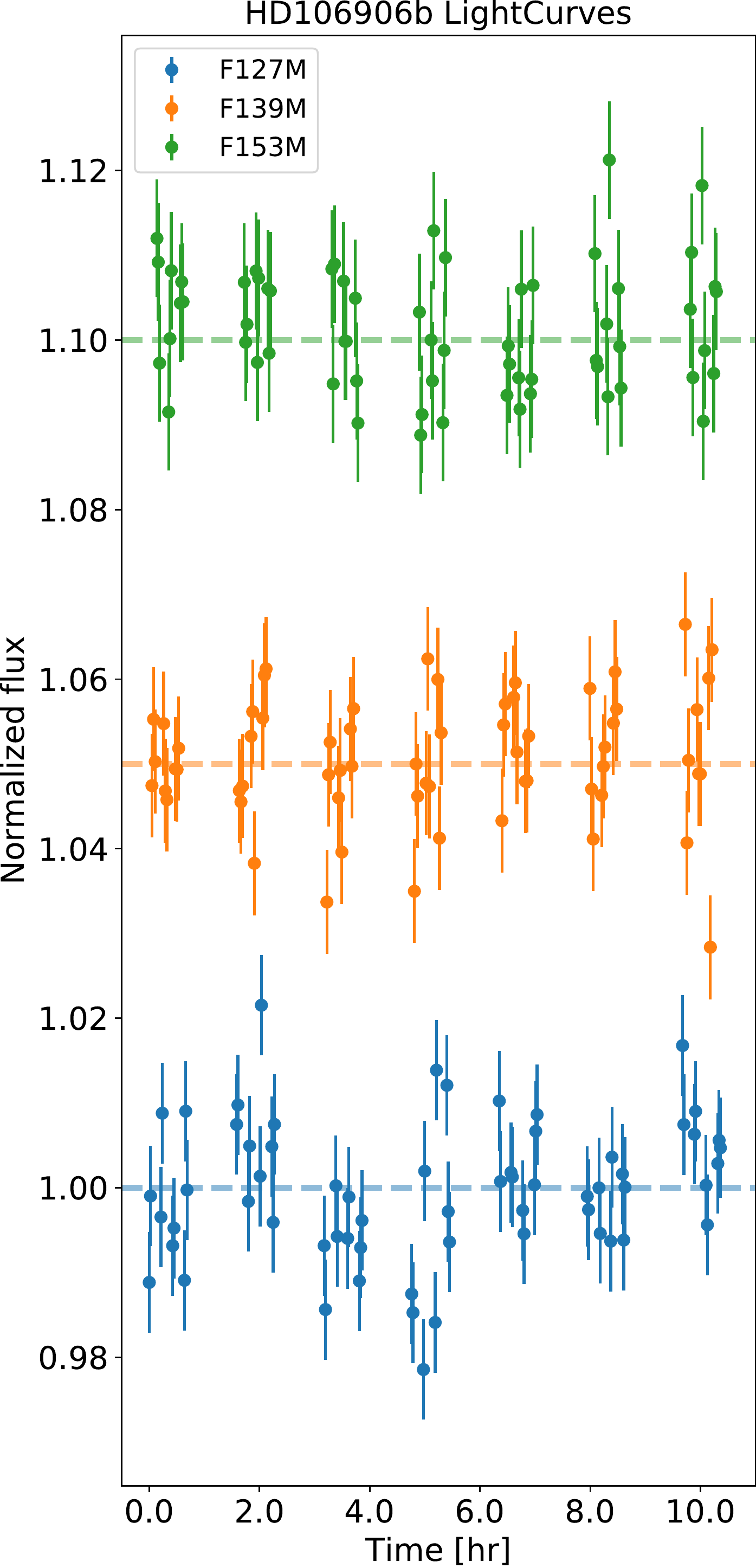}
  \caption{HST/WFC3/IR Lightcurves for HD106906b in the F127M, F139M, and F153M bands. For clarity,  offsets of 5\% and 10\% are applied to the F139M and F153M lightcurves, respectively.}
  \label{fig:lightcurve}
\end{figure}

\subsection{Astrometry}
We follow the procedure detailed in \citet{Bedin2018} for astrometric measurement. Astrometric measurements are made for HD106906 A and b, as well as 25 background stars.
We first measure the raw Cartesian $(x, y)$ coordinates by fitting empirically derived PSFs to the \texttt{flt} images using a software that is adapted from the program \texttt{img2xym\_WFC.09x10}, which is initially developed for ACS/WFC \citep{Anderson2006} and extended for WFC3/IR. The empirical PSFs are from publicly available PSF library\footnote{Released by J. Anderson \url{http://www.stsci.edu/~jayander/WFC3/WFC3IR\_PSFs/}}.  We then apply the most updated geometry correction for WFC3/IR\footnote{Derived by J. Anderson and is publicly available \url{http://www.stsci.edu/~jayander/WFC3/}}. The corrected Cartesian coordinates within the same epoch are then sigma-clipped averaged,  assuming no (sizable) intrinsic motion of sources observed within the same epoch. These procedures result in the geometrically corrected Cartesian coordinates and their uncertainties for each source in each epoch.

We then transform the corrected Cartesian coordinates to the equatorial coordinate system (right ascension, {R.A.}, $\alpha$ and declination, {Dec.}, $\delta$). Common stars with GAIA DR2 astrometry \citep{Gaia2018} are used to find the most general linear transformation (six parameters) that converts $(x, y)$ to $(\xi,\eta)$ (the projections of the equatorial $\alpha$ and $\delta$ coordinates on the tangent plane). $(\xi, \eta)$ are then transformed to $(\alpha, \delta)$ using Equations (3) and (4) in \citet{Bedin2018}.

Considering the non-linearity in $(x, y)$ to $(\alpha, \delta)$ transformation, we adopt a Monte Carlo approach to derive the uncertainties in R.A. and Dec.  For every source, we generate 1,000 Gaussian distributed samples of $(x, y)$  based on the best-fitting values and their uncertainties. We then transform the Cartesian list to a list of R.A. and Dec. pairs. We calculate the standard deviations of the R.A. and Dec. as their $1$-$\sigma$ uncertainties. We note that the uncertainties in R.A. and Dec. include PSF-fitting uncertainties but do not include systematic uncertainties that can be introduced by motions of the reference sources that are used to establish the $(x, y)$ to $(\xi,\eta)$ transformation. i.e., the astrometric measurements and uncertainties are accurate with respect to a single epoch, but the uncertainties may be underestimated for comparison of astrometry between two epochs.

\section{Results and Discussion}
\subsection{Photometry, Lightcurves, and Variability}

Figure~\ref{fig:lightcurve} shows the corrected and normalized lightcurves in the F127M, F139M, and F153M bands. For single exposures, we achieve average photometric signal-to-noise-ratios (SNR) of  77, 78, and 105 in the F127M, F139M, and F153M bands, respectively.
For the lightcurves, variations with zero-to-peak amplitude greater than 1\%  are \emph{not} detected in any bands.
The lightcurve features are dominated by random noise.  Relative to flat lines, the three lightcurves have reduced-$\chi^{2}$ of 1.89, 1.47, and 1.1 in the F127M, F139M, and F153M bands, respectively. Only the F127M lightcurve show a trace of temporal variations while the other two lightcurves fully agree with flat lines. Because we conduct PSF photometry on the uncataloged source that is close to HD106906b, we also obtain its lightcurves. The lightcurves in all three bands of this source are consistent with flat lines and do not show any correlations with lightcurves of HD106906b. The total variations in the lightcurves of this uncataloged source are less than 0.5\% of the fluxes of HD106906b in all three bands. The contamination from the PSF wing of this source to the photometric time-series of HD106906b is thus negligible. Therefore, we can firmly rule out any contaminating signals from this uncataloged source to our variability measurement of HD106906b.

\begin{figure}[!th]
  \centering
  \plotone{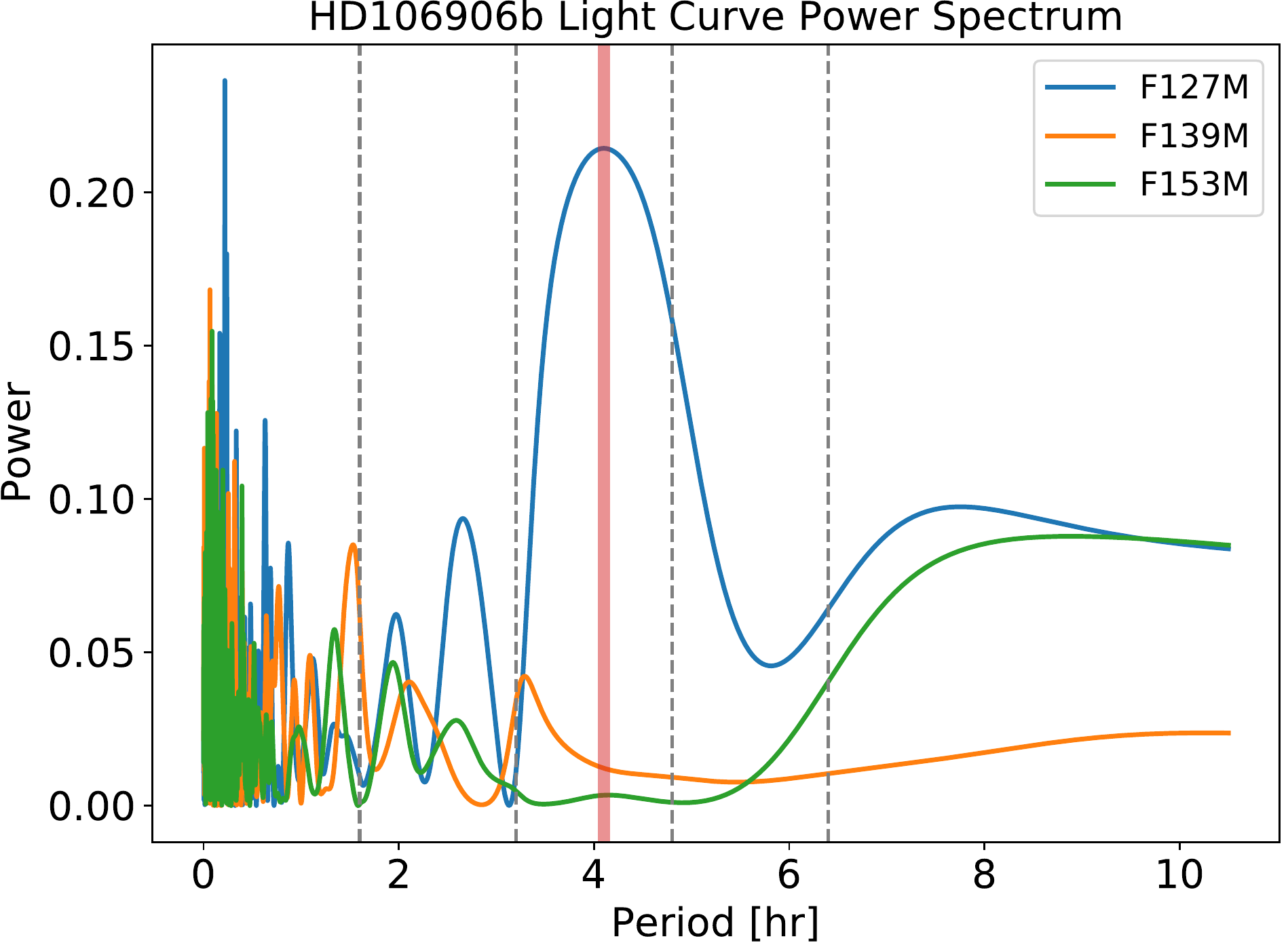}
  \plotone{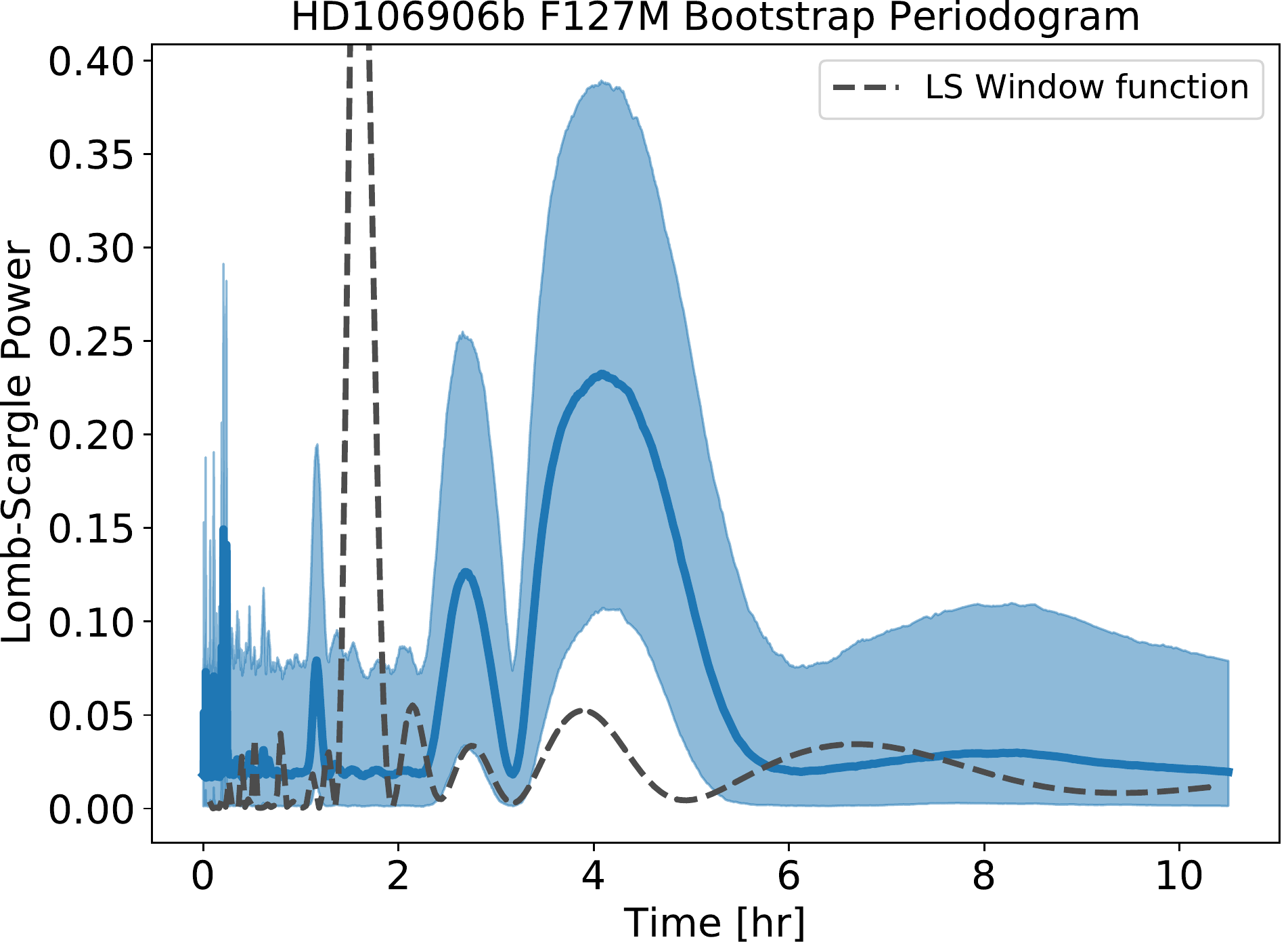}
  \caption{Lomb-Scargle periodogram for the lightcurves of HD106906b. \emph{Upper:} Power spectra for the F127M, F139M, and F153M. The power spectrum for the F127M band lightcurve has a peak at 4~hr. The other two power spectra do not have any significant periodicity detection except the high frequency region dominated by random noise. \emph{Lower:} Significance estimate for the 4~hr signal. Based on a Bootstrap analysis, the significance of the periodic signal in the F127M band lightcurve  is $2.7\sigma$. The black dashed line shows the power spectrum of the observation window function. We note the window function power spectrum, which has its main peak at 1.60~hr (HST's orbital period), also has a side lope at 3.92 hr, close to our 4~hr periodic signal.}
  \label{fig:periodogram}
\end{figure}

\begin{figure}[t]
  \centering
  \includegraphics[width=0.93\columnwidth]{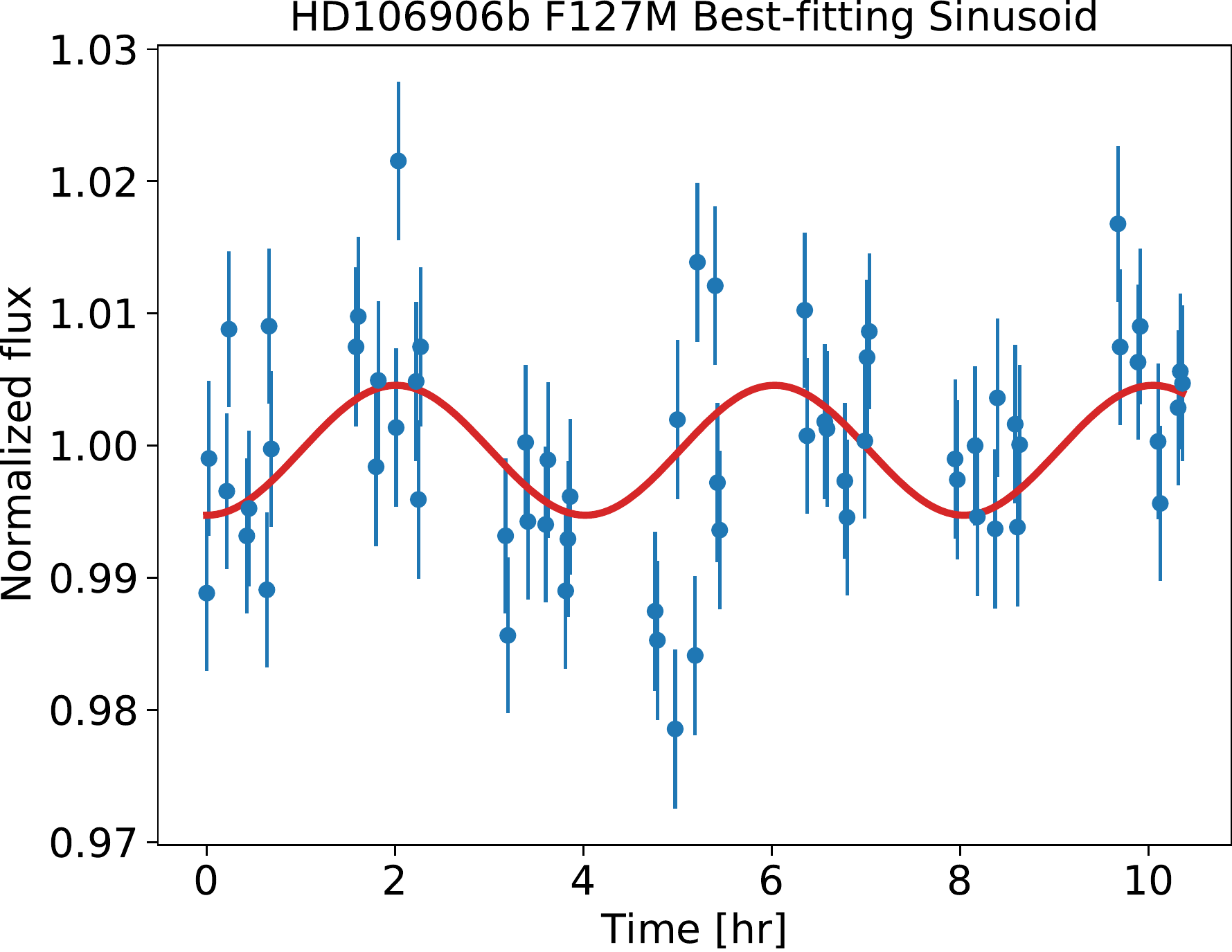}\\
  \includegraphics[width=\columnwidth]{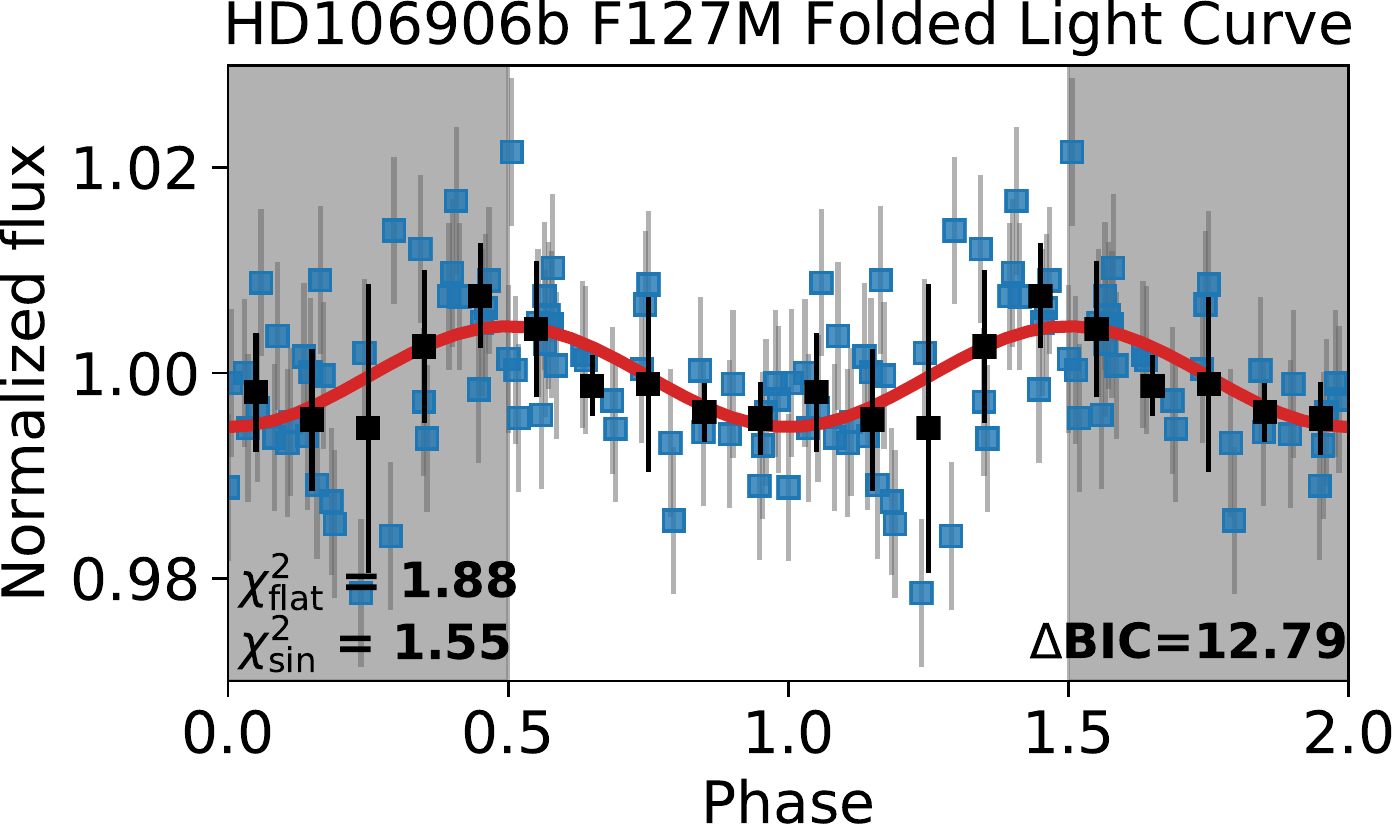}
  \caption{\edit1{HD106906b's lightcurve in F127M and the best-fitting sinusoid. The upper panel shows the original lightcurve and the lower panel shows the lightcurve phase-folded to a period of 4 hr.} This period corresponds to the most significant peak in the Lomb-Scargle periodogram. The red line is the best-fitting sine wave.}
  \label{fig:fold}
\end{figure}

We calculated the Lomb-Scargle power spectra \citep[][Figure~\ref{fig:periodogram}]{Lomb1976} for the lightcurves to investigate lightcurve periodicity. The power spectra for the F139M and F153M lightcurves do not show any significant peaks except in the high-frequency region where the power spectra are dominated by random noise. The lack of signals in the F139M and F153M power spectra is consistent with the featureless lightcurves. The power spectra for the F127M lightcurve has a peak at 4~hr. Compared to a flat line, the best-fitting single sine wave with period fixed at 4~hr marginally decreases the reduced-$\chi^{2}$ from 1.88 to 1.55. For Bayesian Information Criterion \citep[BIC,][]{Schwarz1978}, we find that $\Delta\,\mathrm{BIC}=\mathrm{BIC}_{\mathrm{flat}}-\mathrm{BIC}_{\sin}=12.79$, suggesting the sine wave model is preferred. The best-fitting amplitude of the 4 hr sine wave is $A = 0.49\pm0.12\%$. Figure~\ref{fig:fold} shows the F127M lightcurve folded to the 4 hr period and the best-fitting sine wave. We use a bootstrap method \citep{Manjavacas2017,Zhou2019} to evaluate the significance of the periodogram signal, and show the result in Figure \ref{fig:periodogram}. This analysis yields a $2.7\sigma$ significance of the 4 hr periodic signal. The 4 hr periodic signal also overlaps with a side-lobe of the periodogram of the observation window functions. The low SNR and the effect from observation window function argue \emph{against} 4 hr signal being a robust detection of periodicity in the lightcurve.  In summary, HD106906b only shows marginal  evidence of variability in the F127M band. Lightcurves in the other two bands (water absorption, the red side of water band continuum) are consistent with flat lines.

\subsection{Rotational modulations of HD106906b}
\label{sec:discuss:rotation}

\begin{figure*}[!t]
  \centering
    \plotone{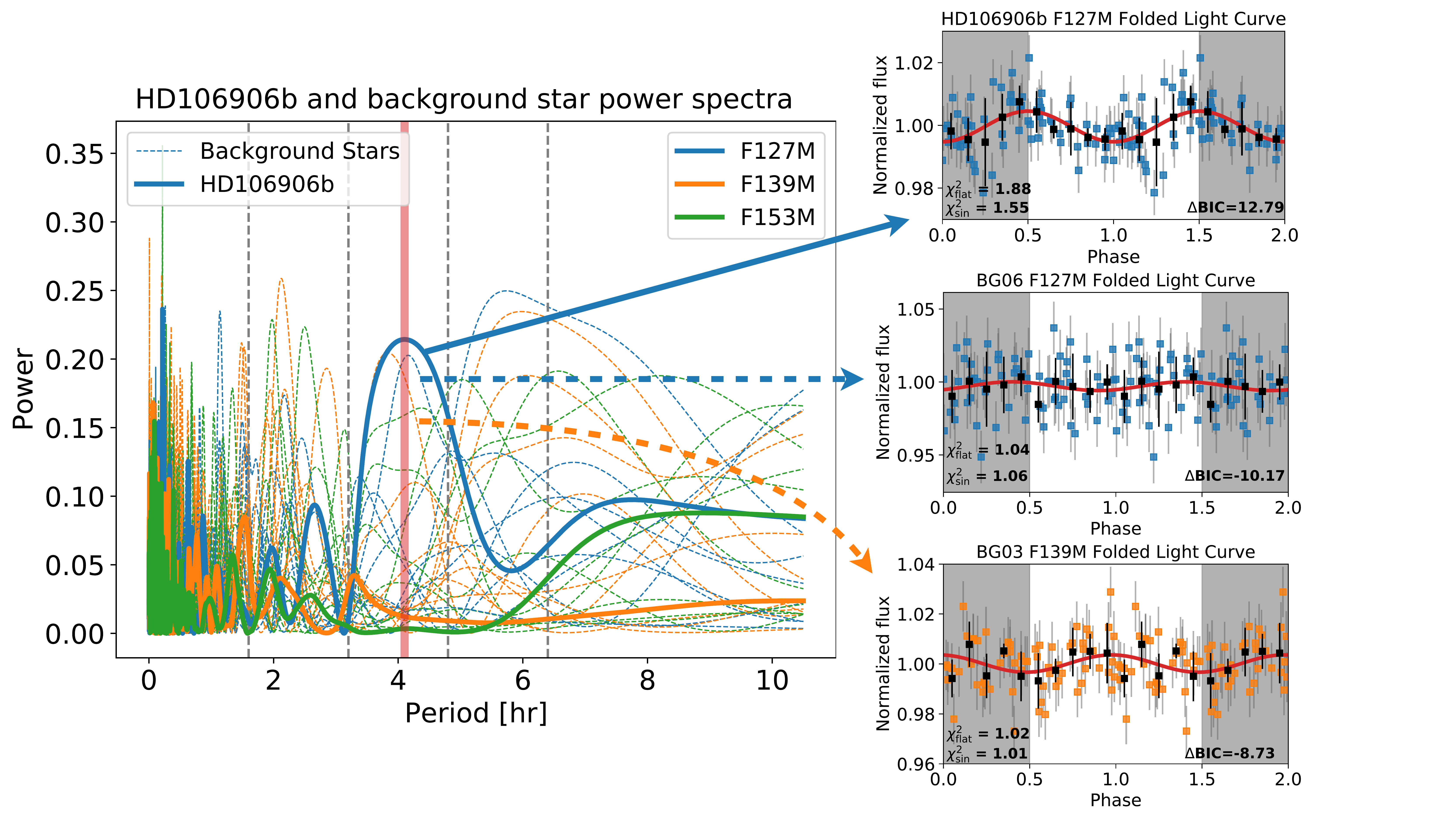}
  \caption{Comparison of the periodograms between those for the ten brightest background stars and those for HD106906b. The two background stars (BG06 in the F127M band and BG03 in the F139M band) that show similar signals to HD109606b's in their periodograms do not show significant variations in the folded lightcurves.}
  \label{fig:all-periodograms}
\end{figure*}

We evaluate the modulation significance in HD106906b's observed lightcurve from both instrumental and astrophysical perspectives. From the instrumental point of view, we have two arguments against the possibility that the modulation signal that we observe in HD106906b's F127M lightcurve arises from instrumental systematics. First, the F127M, F139M, and F153M observations were taken \emph{de facto} contemporaneously with identical instrument set-ups except the choice of filters. Systematics that introduce periodic/sinusoidal signals at 4 hr timescale in the F127M lightcurve should have a similar effect on the other two lightcurves. The agreement of the  F139M and F153M lightcurves with flat lines is inconsistent with the possibility that modulations of the F127M lightcurve {are} due to systematics. Second, similar modulations do not appear in the lightcurves of any of the 20 background stars in the same images. We measure and analyze lightcurves of ten brightest background stars (BG01 to BG10) that are in the field of view of both telescope roll angles and are not affected by the diffraction spikes of the primary PSFs.
Figure \ref{fig:all-periodograms} shows the comparison between the periodograms of the F127M, F139M, and F153M lightcurves of the background stars and that for F127M lightcurve of HD106906b. Most periodograms of the background star do not show significant signals with similar periodicity to HD106906b except two objects (BG03 in the F127M band and BG06 in the F139M band). However, when we fold the lightcurves of those two objects to the periods of the corresponding peaks in the periodograms, the folded lightcurves are consistent with flat lines. For BG03 in the F127M band, the reduced $\chi^{2}$ for a flat line and  the best-fitting sine wave are 1.04 and 1.06, respectively ($\Delta\,\mathrm{BIC}=-10.17$). For BG06 in the F139M band, the reduced $\chi^{2}$ for a flat line and the best-fitting sine wave are 1.02 and 1.01,  respectively ($\Delta\,\mathrm{BIC}=-8.73$). Flat lines are favored in both background star lightcurves, which is opposite to the case for the F127M lightcurve of HD106906b.

From the astrophysical perspective, we can qualitatively evaluate the likelihood for HD106906b, an early L-type planetary-mass companion to be rotationally modulated only in the F127M band but not in the other two bands. Multi-wavelength and time-resolved observations of ultra-cool dwarfs have found that the rotational modulations for the majority  brown dwarfs and planetary-mass companions are wavelength-dependent and have higher amplitudes at shorter wavelengths than longer wavelengths \citep[e.g.,][]{Apai2013,Yang2015,Zhou2016,Schlawin2017,Zhou2019}. These findings are consistent with a model prediction based on Mie-scattering calculation \citep{Hiranaka2016,Lew2016,Schlawin2017}. Additionally, the 1.4 \micron{} water absorption or the F139M band sometimes show reduced rotational modulation amplitude \citep[e.g.,][]{Apai2013}, due to water vapor opacity elevating the photosphere at this wavelength.  Therefore, rotational modulations only appearing in the band with the shortest wavelength of our observation is qualitatively consistent with model predictions and previous observations, particularly those for planetary-mass companions \citep{Zhou2016,Zhou2019}. If we assume that the wavelength dependence of HD106906b's rotational modulations is the same as that measured in 2M1207b \citep{Zhou2016} as 2M1207b (a mid-L-type planetary-mass companion) is HD106906b's close analog that also has modulation detected, we would expect the modulation amplitude in the F153M band to be 0.6\%. Our observation is not sensitive to such small amplitude modulations. Therefore, if the overall modulation amplitude is low, it is likely that the signal is only detected in the bluest band of the observation, which is consistent with our observations.

These two lines of evidence support the interpretation that the modulations we see in HD106906b's F127M lightcurve are astrophysical and, in particular, caused by heterogeneous clouds. Nevertheless, we emphasize that the amplitude of the signal is marginal. Our evaluation of the rotational modulation and rotation period for HD106906b remain \emph{tentative}.  \edit1{The lack of large-amplitude rotational modulations in the HD 106906b lightcurve might be indicative of a (nearly) pole-on geometry of its rotational axis \citep[e.g.,][]{Vos2017b}. 
  This prediction can be tested by $v\sin i$ measurements from high-resolution spectroscopic observations \citep[e.g.,][]{Snellen2014,Vos2017b,Bryan2018}.}

Applying the rotational break-up limit criterion provided in \citet{Marley2011}, which is a function of radius and surface gravity, we find that HD106906b will break up if its rotational period is shorter than 1.44 hr. The rotation rate that corresponds to a 4 hr period is significantly below this limit.

\subsection{Spectral Energy Distribution}

\begin{figure*}
  \centering
  \plottwo{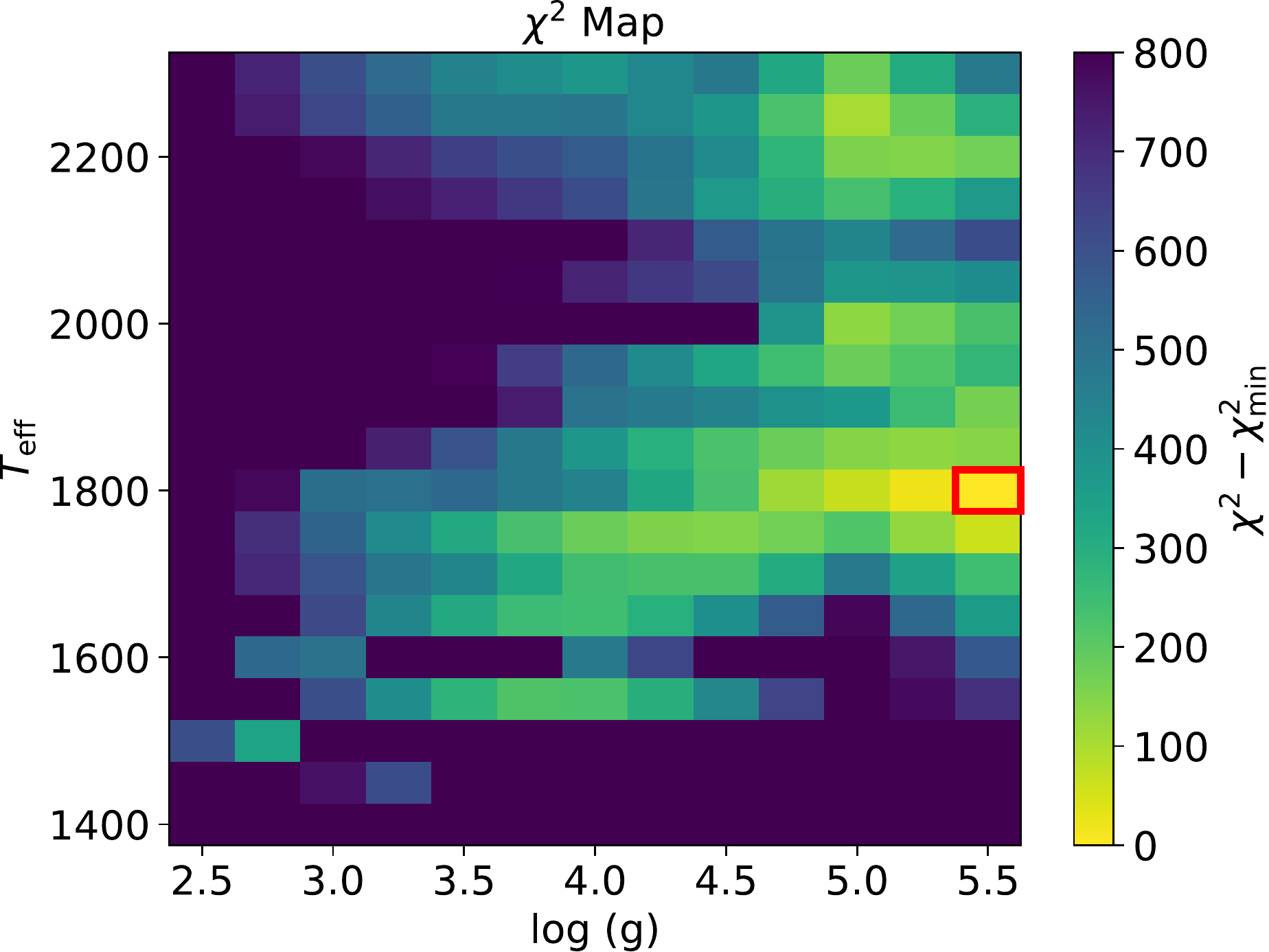}{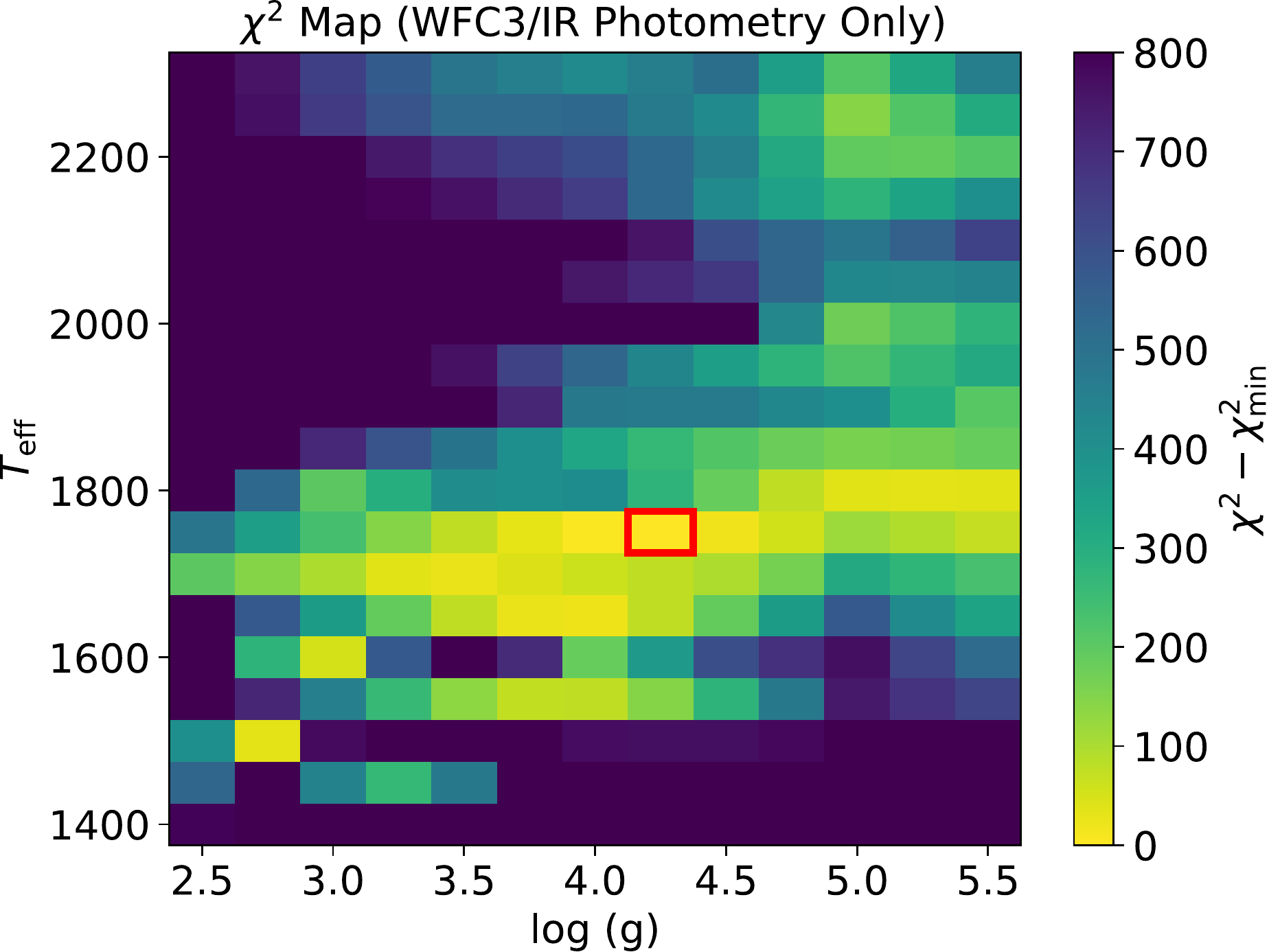}
  \plottwo{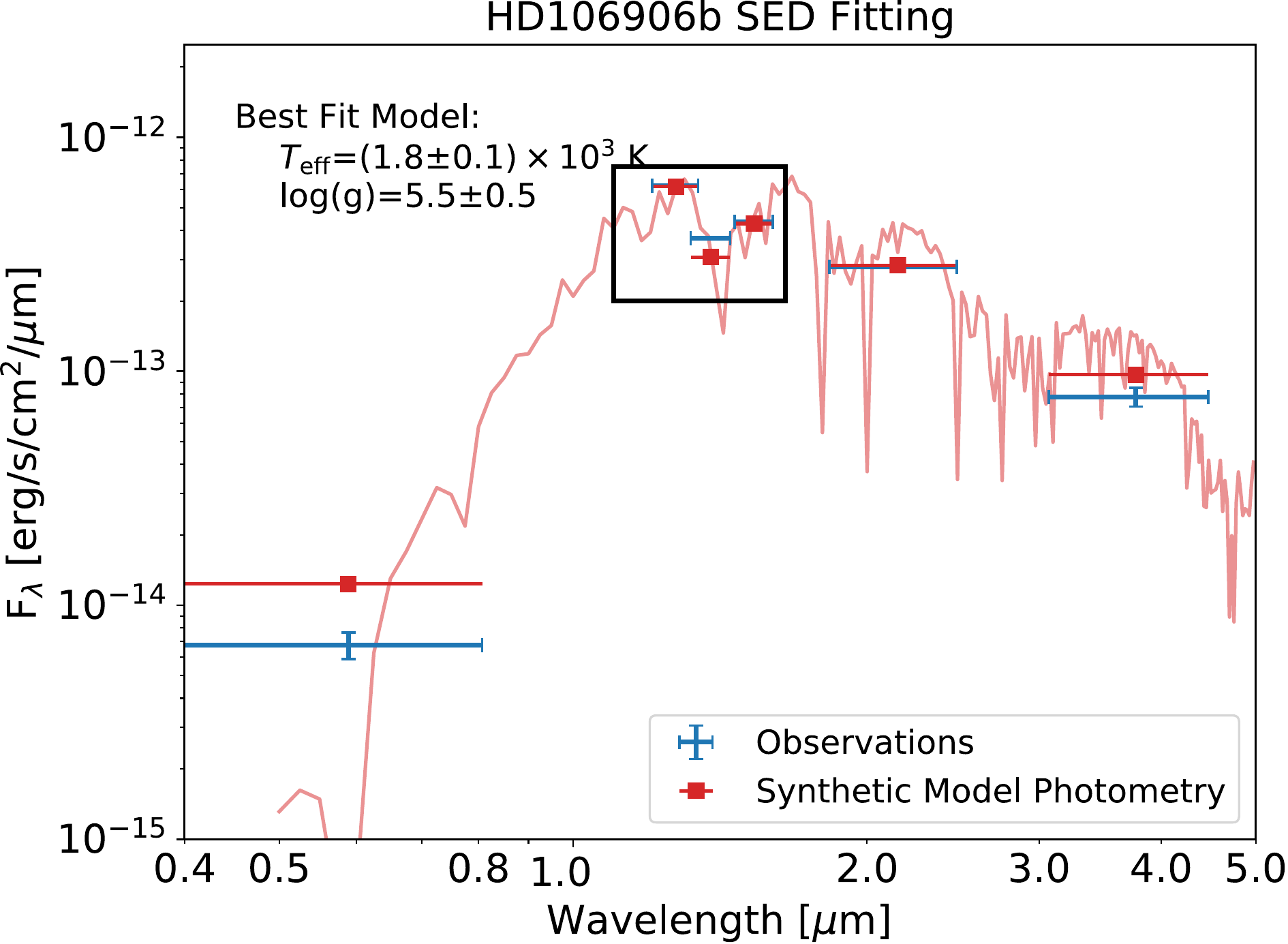}{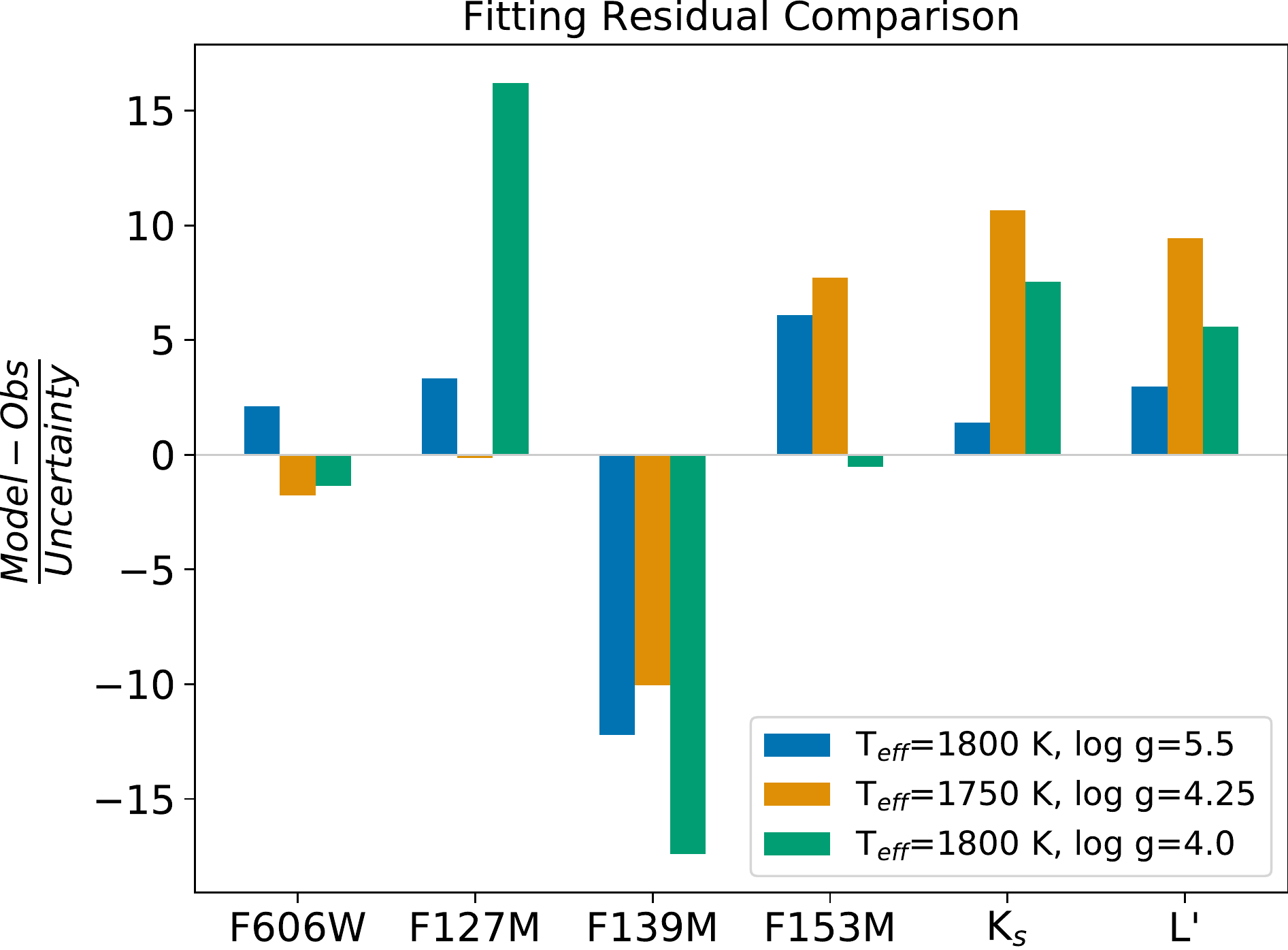}
  \caption{The comparison of the SED of HD106906b to the BT-Settl models. The upper panel shows the $\chi^{2}$ as a function of model \teff{} and \logg{} for a fit that includes all six bands (left) and a fit that only includes the three WFC3/IR bands (right). The grid points that yield the lowest $\chi^{2}$ are marked by red rectangles (\teff{}=1,800 K and \logg{}=5.5 for the full fit and \teff{}=1,750 K and \logg{}=4.25 for the WFC3/IR-only fit). The lower left panel shows the full observed SED (blue dot) and the best-fitting (1800~K, $\log g=5.5$) BT-Settl spectrum (red line) and the synthetic model photometry (red square). The lower right panel compares the fitting residuals for the best-fitting model (\teff{}=1,800 K, \logg=5.5), the model best-fit to the WFC3/IR (\teff{}=1,750, \logg=4.25), and an intermediate gravity model (\teff{}=1,800 K, \logg=4.0).}
  \label{fig:SED}
\end{figure*}

Our precise time-averaged photometry, particularly HD\-106906b's flux density in the water absorption band is useful for determining fundamental properties, such as \teff{} and \logg{} of HD106906b through spectral energy distribution (SED) fitting.  We combine our photometry with archival photometry to form the SED of HD106906b.  We use HST/ACS/F606W band photometry ($\lambda_{\mathrm{pivot}}=0.596\micron$, FWHM$=0.234\micron$) from \citet{Kalas2015}, $K_{s}$ ($\lambda_{\mathrm{pivot}}=2.145\micron$, FWHM$=0.305\micron$) and $L'$ ($\lambda_{\mathrm{pivot}}=3.774\micron$, FWHM$=0.592\micron$) band photometry from \citet{Bailey2013}. We do not use the archival $J$ band photometry because our F127M photometry covers similar spectral features and has more than $20\times$ greater SNR. Importantly, our F139M photometry provides a tight $1.4\micron$ water absorption constraint for HD106906b.

We fit the SED of HD106906b to the BT-Settl model grid \citep[][]{Allard2012} and present the results in Figure~\ref{fig:SED}. To account for filter throughput and the target's flux density variation within each band, we use \texttt{pysynphot}\footnote{https://pysynphot.readthedocs.io/en/latest/} to convert the model spectrum to flux density in count rates for the three WFC3/IR filter bands. For the archival photometry, which are presented in AB magnitude (HST/ACS/F606W) and in Vega magnitude ($K_{s}$ and $L'$), the BT-Settl model are directly available in the corresponding magnitude systems.  For SED fitting, we bi-linearly (in \teff and \logg dimensions) interpolate the model grid (native grid resolution: $\Delta \teff=100\,\mbox{K}, \Delta \log g=0.5$) in magnitude scales. The free parameters are effective temperature $\teff$, surface gravity $\logg$, and scaling parameter $\mathcal{S}$, the ratio between the observed flux over model flux. Because model SEDs are presented in flux {density} at the photosphere surface, the scaling parameter can be transformed to the photospheric radius via $R=\sqrt{\mathcal{S}}\,d$, in which $d$ is the distance of the system. By searching for the minimum $\chi^{2}$ \edit1{(the upper panel of Figure \ref{fig:SED})}, we identify the best-fitting $T_{\mathrm{eff}}=1,800\pm100$~K and $\log g=5.5\pm0.5$.  The scaling parameter corresponds to a radius of  $1.775\pm0.015R_{\mathrm{Jup}}$ at a distance of 103.3~pc \citep{Gaia2018,Gaia2016}. The 1,800~K effective temperature estimate is consistent with previous studies \citep{Bailey2013,Wu2016}, but the surface gravity is not compatible with a low surface gravity assessment.

To investigate the fitting results, we further examine the SED fitting residuals. As demonstrated in the lower right panel of Figure~\ref{fig:SED}, although the $T_{\mathrm{eff}}=1,800 \mathrm{K},\log g=5.5$ model reproduces the overall shape of HD106906b's SED, it under-predicts the flux density in the F139M band (i.e., over-predict the 1.4\micron{} water absorption depth). Because the photometric measurements in the three WFC3/IR bands have the smallest uncertainties (~1\%), they have the largest contributions to the $\chi^{2}$ statistics. Thus the mismatch in the F139M band is more than $10\sigma$ leading to a $\chi^{2}>100$ (degrees of freedom=3) even for the best-fitting model. With an intermediate gravity model with the same temperature ($T_{\mathrm{eff}}=1,800 \mathrm{K},\log g=4.0$), the disagreement between observations and the model at the 1.4\micron{} water band is more prominent, causing the $\chi^{2}$ statistics to increase by more than 300. Thus the high-gravity model is favored. Considering the strong diagnostic power of the precise WFC3/IR measurements, we conduct an additional fit that only includes photometry in those band. When the constraints from the longer wavelength ($K_{s}$ and $L'$ bands) are ignored, the SED fitting demonstrates a \teff{}-\logg{} degeneracy in the \teff{} range of $1,600$~K to $1,800$~K and favors a slightly cooler and intermediate gravity model ($T_{\mathrm{eff}}=1,800 \mathrm{K},\log g=4.0$). However, comparing to the complete SED, this model does not reproduce the overall shape and is thus disfavored in the full SED fit. In summary, although the best-fitting parameters are robust in our least-$\chi^{2}$ fitting framework, because of the large residuals at the 1.4 \micron{} water absorption band, this result should not be taken as evidence for high surface gravity of HD106906b, but a demonstration of challenges in modeling the spectra of ultracool atmospheres of young planetary-mass objects.

\subsection{Astrometry}
\label{sec:astrometry}

In order to establish a precise astrometric reference frame and constrain the relative motion between HD106906b and its host star, we measure the R.A. and Dec. of 25 sources (BG01 to BG25) that are in the field of view (FoV) of both HST/WFC3 epochs. The average uncertainties in R.A. and Dec. are 5.3 mas for the 2016 epoch and 2.9 mas for the 2018 epoch, corresponding to 0.041 and 0.023 pixels, respectively. Due to the saturation at the PSF core, HD106906A has one of the {lowest} astrometric precisions of all the sources. Especially in the 2016 epoch, its astrometric uncertainty is 51.2 mas or 0.39 pixel. Astrometric measurements for HD106906 are listed in Table \ref{tab:astrometry} and those for the background sources are listed in Table \ref{tab:bck} in the appendix.

We derive the separations and position angles between HD106906A and b and their uncertainties for the 2016 and 2018 epochs. The separations are $7.11\arcsec\pm0.03\arcsec$ and $7.108\arcsec\pm0.005\arcsec$ in the 2016 and 2018 epochs, respectively. The position angles are $307.5^{\circ}\pm0.3^{\circ}$ and $307.29\pm0.05^{{\circ}}$ in the two epochs, respectively. These separations and position angles are indistinguishable from those measured in the ACS/HRC images \citep{Bailey2013}. Therefore, relative motions between the companion and the star are not detected. The substantial positional uncertainty of HD106906A due to saturation is the bottleneck that limits the astrometric value of these HST images.

\begin{deluxetable}{llclc}
  
  \tablecaption{HST/WFC3 Astrometry for HD106906 System.\label{tab:astrometry}}
  
  \tablehead{
    \colhead{Object (epoch)} &
    \colhead{{R.A.}} &
    \colhead{{R.A.}$_{\mathrm{err}}$} &
    \colhead{{Dec.}} &
    \colhead{{Dec.}$_{\mathrm{err}}$}\\
    \colhead{} &
    \colhead{[hh mm ss]} &
    \colhead{[mas]} &
    \colhead{[dd mm ss]} &
    \colhead{[mas]} 
  }
  
  \startdata
  HD106906A (2016) & 12 17 53.118 & 16 & $-$55 58 32.136 & 49 \\
  HD106906b (2016) & 12 17 52.444 & 1.1 & $-$55 58 27.8199 & 0.79 \\
  HD106906A (2018) & 12 17 53.108 & 5.6 & $-$55 58 32.158 & 6.7 \\
  HD106906b (2018) & 12 17 52.434 & 2.1 & $-$55 58 27.843 & 2.3 \\
  \enddata
\end{deluxetable}

With a temporal baseline of 14 years, three epochs of HST observations are not able to detect relative motion between HD106906b and its host star. Assuming a face-on, circular orbit and an orbital radius of 732 au, we expect an orbital arc length for HD106906b to be 37.1~mas in 14 yr (first epoch with ACS in 2004) or 5~mas in 2~yr (between the two WFC3 epochs). These arc lengths correspond to $12.8\times$ and $1.72\times$ the average $1\sigma$ astrometric uncertainty in the 2018 epoch. As a result, the HST images could resolve the first orbital motion if their precisions are not limited by saturation.

Astrometric constraints of the HD106906 system are critical to studying the system's formation and dynamical evolution history \citep[e.g., ][]{DeRosa2019} and for measuring the dynamical mass of the planet \citep[e.g.,][]{Snellen2018,Dupuy2019}. The design of future observations should consider optimization for astrometric precisions, which includes avoiding saturation, increasing spatial resolution through dithering, repeating at the same celestial orientation angles and re-use of the same guide stars. In the $33\times33$ arcsec$^{2}$ FoV of WFC3 images, there are seven background sources that have {celestial} coordinates and proper motion measurements from GAIA DR2. Using these sources to register the WFC3 image with GAIA can calibrate the absolute astrometry to sub-mas precision level \citep{Bedin2018}. Future astrometric analysis of HD106906 system will  benefit from our background source catalog (Table \ref{tab:bck}).

\subsection{Other Sources in the Field of View}
\label{sec:othersources}

\begin{figure*}[!t]
  \centering
  \includegraphics[width=0.32\textwidth]{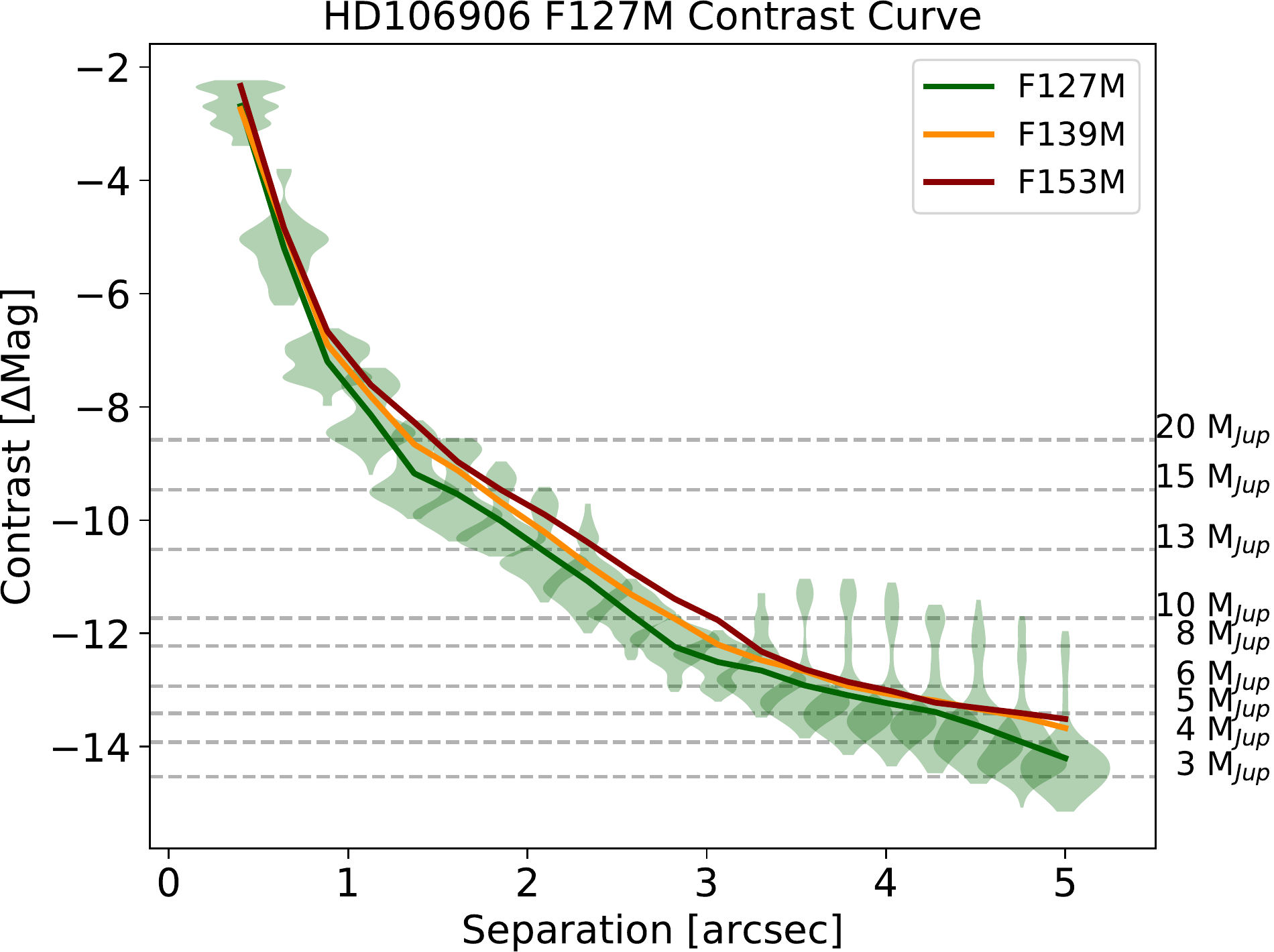}
  \includegraphics[width=0.32\textwidth]{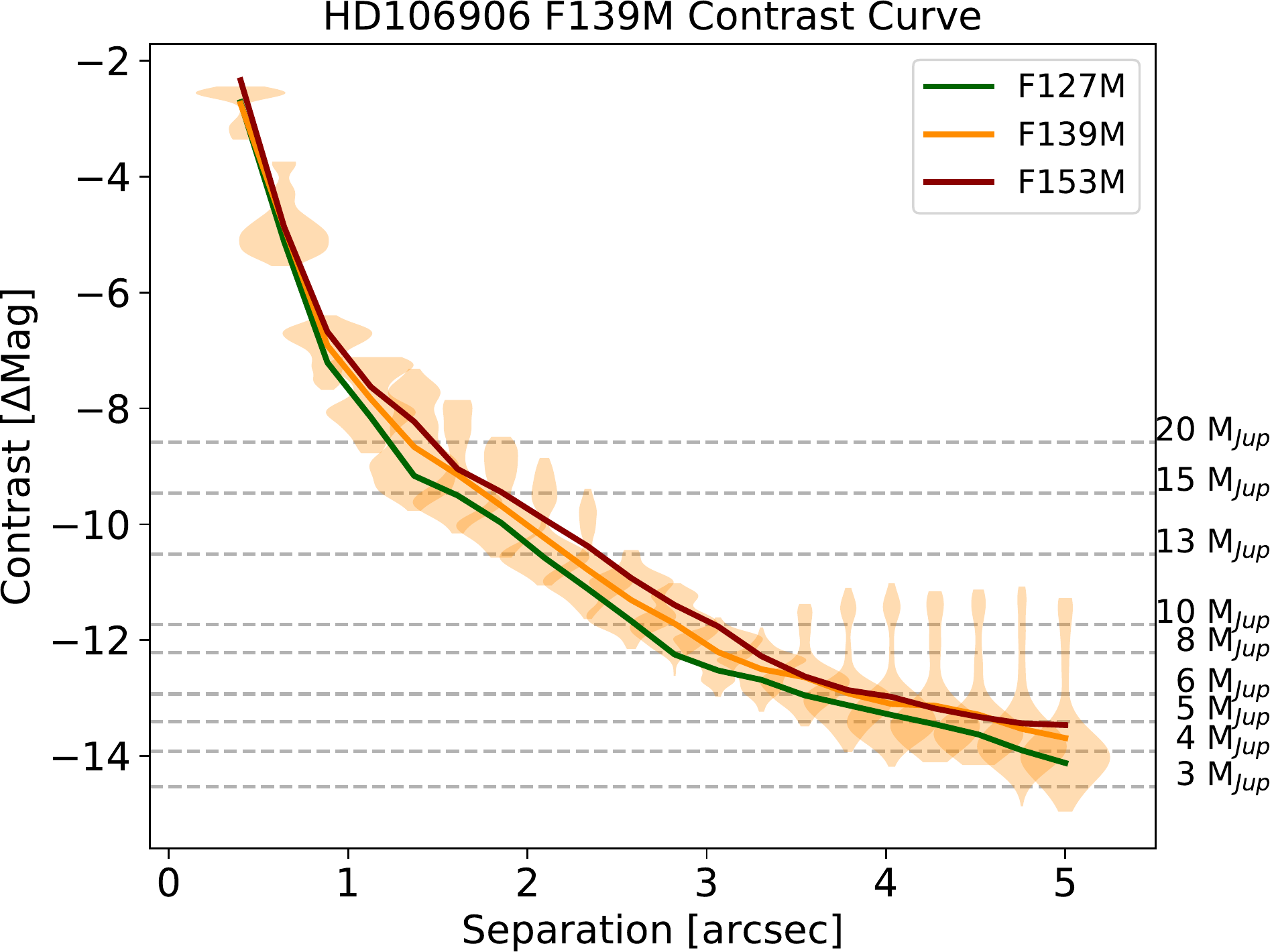}
  \includegraphics[width=0.32\textwidth]{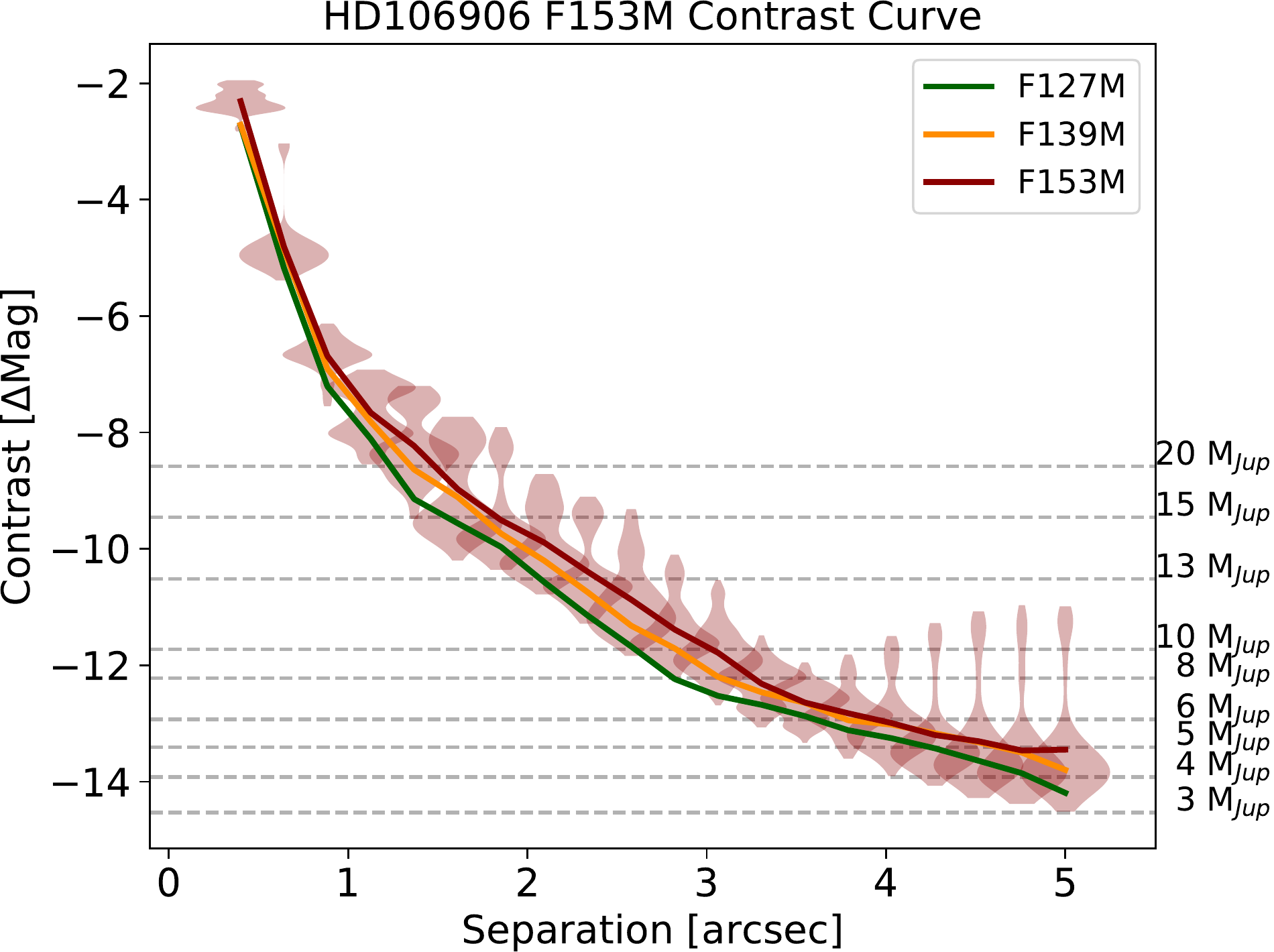}
  \caption{Azimuthally averaged contrast curves in the F127M, F139M, and F153M band images for the HD106906 observations. Violin plots are adopted to demonstrate the contrast distribution at a certain separation. Because of the spatial variance in the primary subtracted image, the contrast does not fully follow a Gaussian distribution. Each panel shows the contrast distributions in one of the three bands and the averaged contrast curves in all three bands are presented in every panel. The corresponding companion mass estimates from an evolutionary model \citep[$f_{\mathrm{sed}}=2$ cloud, 15 Myr]{Saumon2008} are annotated on the right side of each plot.}
  \label{fig:contrast_curve}
\end{figure*}

In order to assess the possible presence of yet undetected companions to HD106906, we construct $33\arcsec\times33\arcsec$ FoV deep images (Figure \ref{fig:bck}) by median-combining the entire HST/WFC3/IR time series for each filter. These images may include yet undiscovered companions of HD106906A. With our observational setup, the water absorption depth can be an effective criterion for selecting candidate ultra-cool objects \citep[e.g.,][]{Fontanive2018}. Here we define the absolute water absorption depth as the difference between the F139M flux density and the average flux {density} in the F127M and F153M bands. We further define the normalized water absorption depth ($\mathcal{D}$) as the absolute depth divided by the average flux density in the F127M and the F153M bands. $\mathcal{D}$ is calculated as

\begin{equation}
  \label{eqn:water}
\mathcal{D} = \frac{(f_{\mathrm{F127M}} + f_{\mathrm{F153M}})/2 - f_{\mathrm{F139M}}}{(f_{\mathrm{F127M}} + f_{\mathrm{F153M}})/2} 
\end{equation}

In all three bands, we calculate the $5\sigma$ contrast curves for contrast-limited point-source detections for the median-combined primary-subtracted images (Figure \ref{fig:contrast_curve}). We construct these contrast curves through a PSF injection-and-recovery process, as detailed in  \citet{Zhou2019}.  We find that the three bands  have almost identical contrast curves, although the F127M image has the deepest contrast at wide separation.  Our median-combined, primary-subtracted images are sensitive to $\Delta \mbox{mag}=7.7$ at 1\arcsec, $\Delta \mbox{mag}=10.4$ at 2\arcsec, and $\Delta \mbox{mag}=14.2$ at 5\arcsec. Assuming an age of 15 Myr and the evolution tracks of \citet{Saumon2008} ($f_{\mathrm{sed}}=2$ cloud), our median-combined, primary-subtracted images can place 5$\sigma$ upper limits for companions more massive than 13\mjup{} at 2\arcsec{} or greater  separations and 4\mjup{} at 4.75\arcsec{} or greater  separations.

\begin{figure}
  \centering
  \plotone{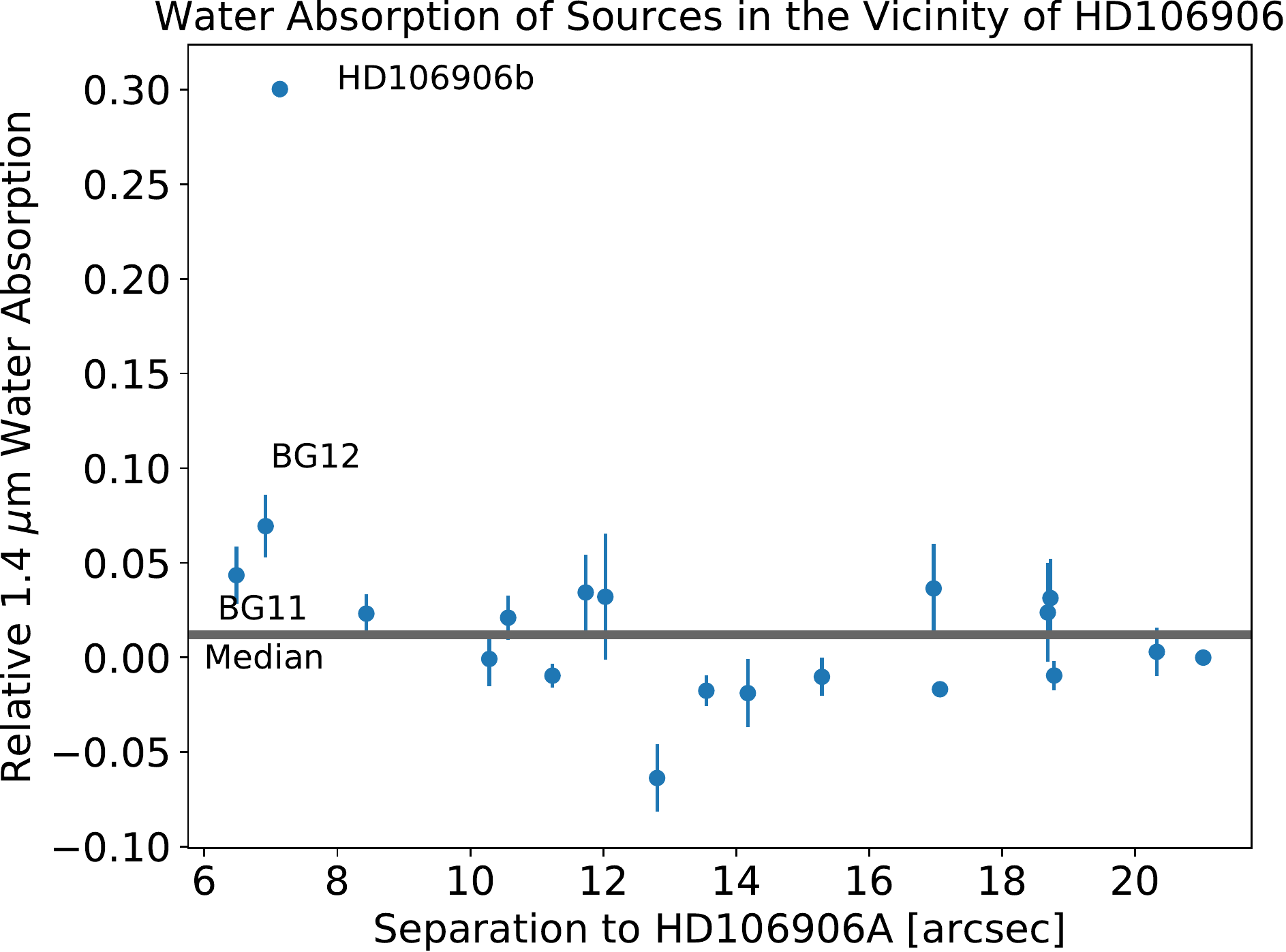}
  \caption{Measured relative water absorption depths of 25 sources in the field of view. The sources are ranked by their angular distance to HD106906A. Except HD106906b, there are two sources (BG11 and BG12) have water absorptions, but at much weaker levels.}
  \label{fig:backgroundsources}
\end{figure}

We used the median-combined primary-subtracted images to measure the relative water absorption depth for 25 point sources (from BG01 to BG25, see Table \ref{tab:bck}) that are in the field of view for images taken with both telescope rolls. Figure~\ref{fig:backgroundsources} shows the water absorption depth for each source.  Water absorption is marginally detected in two other sources (BG11 and BG12). Interestingly, these two sources also have the smallest angular separations from HD106906A among all point sources in the field of view. For both BG11 and BG12, their astrometry in the 2016 and 2018 HST/WFC3 observations are consistent within 15~mas and they do not appear to co-move with the HD106906 system. Therefore they are likely background stars.  BG12 {is also very close to HD106906b in angular separation} (0.87\arcsec) and is also in the field of view of the 2004 HST/ACS image.   Based on the HST/ACS/HRC and the HST/WFC3/IR photometry, the SED of this source is fully consistent with a $3,700\pm100$~K BT-Settl stellar SED model. BG12's fourteen-year baseline astrometry is consistent with that for a stationary background source. Therefore, BG12 is most likely a background K/M giant star. 

\begin{figure}[!t]
  \centering
  \plotone{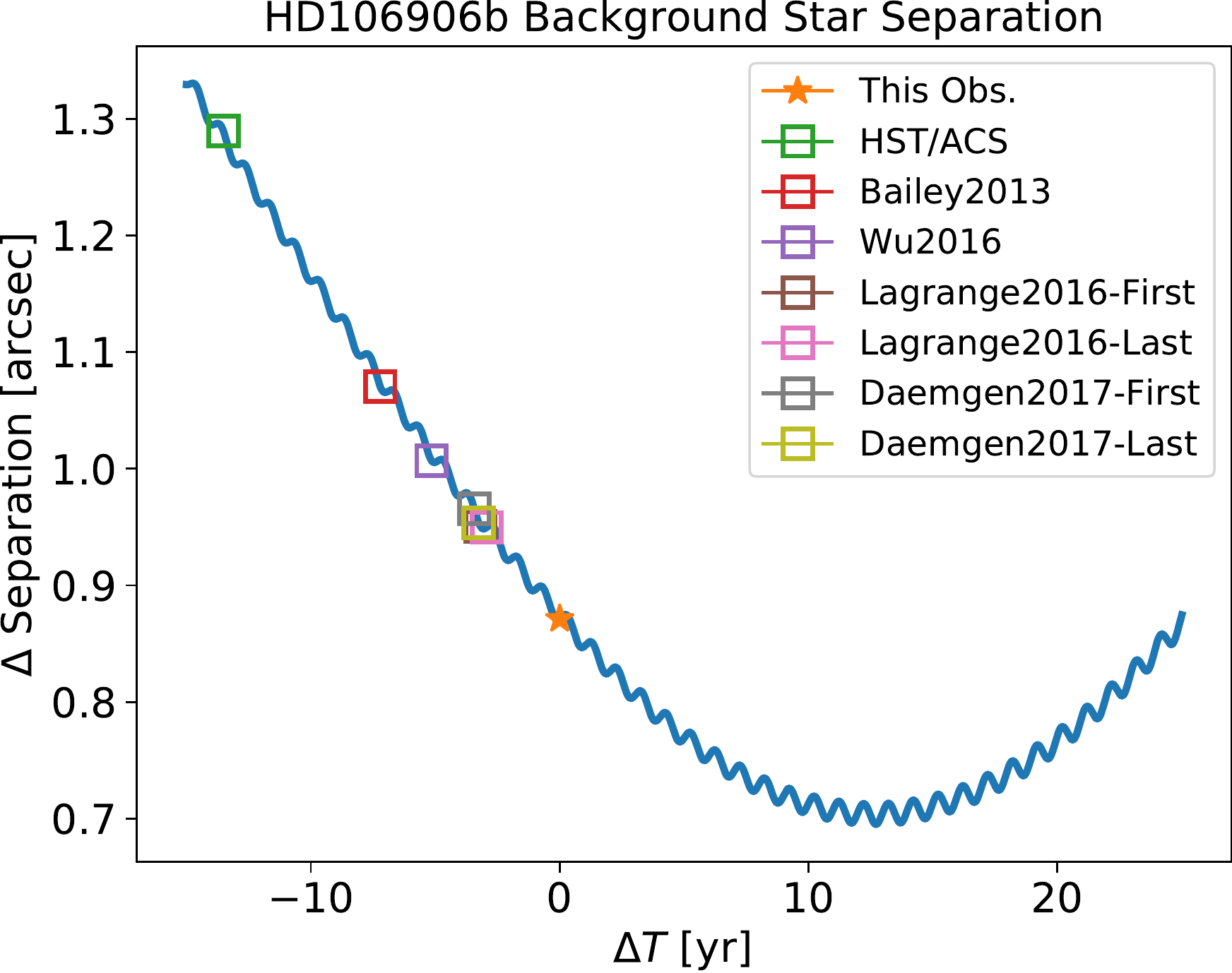}
  \caption{Simulated separations between HD106906b and BG12, assuming it is a stationary background star. Under this assumption, HD106906b and BG12 will reach their minimum separation at 0.695\arcsec{} in 2031. The predicted separations in the past observations of HD106906b are marked by squares.}
  \label{fig:astrometry:bck}
\end{figure}

We investigate the apparent trajectory of the background source BG12, noting that its location at prior epochs could potentially have contaminated observations of HD106906b reported earlier in the literature.  We calculate  the differences in right ascension ($\Delta${R.A.}),  declination ($\Delta${Dec.}), and the separations between HD106906b and the close background source from the year 2003 (one year before the first direct imaging {reported for} HD106906b) to the year 2023. In this calculation, the close background source is assumed to be stationary and HD106906b is co-moving with its host star at $(\mu_\alpha\cos\delta=-39.01\,\mbox{mas/yr}, \mu_{\delta}=-12.87\,\mbox{mas/yr})$ \citep{Gaia2016, Gaia2018}. The results are shown in Figure~\ref{fig:astrometry:bck}. In the same figure, we also marked the expected positions of the close companion in previous observations \citep{Bailey2013, Wu2016, Lagrange2016, Daemgen2017} assuming BG12 is a stationary background star.

Figure~\ref{fig:astrometry:bck} demonstrates that HD106906b, due to its proper motion, has been approaching  -- in projection -- to {the location of BG12}  over the years. The separation between HD1006906b and {BG12} has been decreasing from 1.29\arcsec (2004, first available image) to 0.87\arcsec (this study), and will reach its minimum at 0.695\arcsec in 2031. In the study of \citep{Bailey2013, Wu2016, Daemgen2017},  HD106906b should have a separation of 0.95\arcsec{}-1.05\arcsec to {BG12}, assuming it is stationary. It is unlikely that {BG12} contaminated those measurements, because the separations in those observation epochs were significantly greater than the spatial resolutions of those observations. Considering the brightness contrast of the two objects, in the worst case in which BG12 is entirely included in the aperture for measurements of HD106906b, the contamination of the background star to HD106906b's broadband photometry is  $<7.5\%$ in the near-infrared.

\section{Summary and Conclusions}

\begin{enumerate}[labelsep=0.5em, leftmargin=1em]
\item We observed the planetary-mass companion HD106906b with seven consecutive HST orbits in HST WFC3/IR's direct-imaging mode. Applying two-roll differential imaging and PSF-fitting photometry, we have achieved single-frame photometric precisions of 1.3\%, 1.3\% and 0.9\% in the lightcurves in the F127M, F139M, and F153M bands, respectively. The F127M lightcurve shows a tentative ($2.7\sigma$) variability signal that best-fit by a $P=4$\,hr rotational modulation.
\item The marginal detection of the F127M band modulation and the non-detections in the other two bands are consistent with the wavelength dependence of modulation amplitudes previously observed in other brown dwarfs and planetary-mass companions. The marginally-detected, low-amplitude modulations agree with the expectation that early-L type dwarfs are less likely to be have large-amplitude variability compared to the L/T transition types \citep[e.g.,][]{Radigan2014,Metchev2015}. However, due to the low detection significance, the modulation signal cannot serve as conclusive evidence for heterogeneous clouds in the atmospheres of HD106906b.
\item Our observations provide precise  photometry for HD106906b in the HST/WFC3/IR F127M, F139M, and F153M bands. This is also the first precision photometric measurement for HD106906b in the 1.4\,\micron{} water absorption band. We combine our three bands of photometry with archival data to form an SED for HD106906b and perform SED model fitting on the BT-Settl model grid. We find a best-fitting effective temperature of 1800~K, consistent with literature results, and a best-fitting surface gravity $\log g$ of 5.5,  significantly higher than previous estimates and inconsistent with HD106906b being a young and planetary mass object. Also, the observed F139M band flux intensity for HD106906b is significantly higher than the best-fitting model value. Considering the large residuals even in the best-fit model, this finding should not be taken as conclusive evidence of a high surface gravity for HD106906b but rather an indication of the challenges in SED modeling of ultra-cool atmospheres.
\item We combine WFC3/IR images to form primary-subtracted deep images and search for planetary-mass companions in the field of view. Our composite images are sensitive to planets with masses down to $4\mjup$. We used measurements of the 1.4\micron{} water absorption to {arbitrate between close companion candidates and background stars} (i.e., substellar companions should show significant water absorptions). We did not discover  any new companions. We did find two point sources that have lower fluxes in the F139M band. However, both sources do not appear to co-move with the HD106906 system. One of the two objects is in close vicinity to HD106906b (0.85\arcsec angular separation). Based on its astrometry and SED fitting results, this object is likely a background K/M giant star.  Based on GAIA DR2 astrometry and proper motion, the angular distance between HD106906b and this background star is decreasing and will be on the level of 0.7\arcsec to 0.8\arcsec{}  in the 2020s. Future observations of HD106906b will need to  carefully eliminate the flux contamination from this background star.
\item We measured astrometry for HD106906A and b, as well as for the background sources. The separations and position angles between HD106906A and b in the 2016 and 2018 epochs WFC3 images do not deviate from  those in the 2004 ACS/HRC images for more than 1$\sigma$ uncertainty. The saturated PSF core of  HD106906A limits our sensitivity in probing the relative motion between HD106906A and b. HST/WFC3 observations that avoid saturating the primary will at least place strong constraint on whether HD106906b is on a face-on circular orbit and may even resolve the planet's orbital motions.
\end{enumerate}

\software{Numpy\&Scipy \citep{VanderWalt2011}, Matplotlib
  \citep{Hunter2007}, IPython \citep{Perez2007}, Astropy
  \citep{Robitaille2013}, Seaborn \citep{Waskom2017}, Image Registration \citep{Ginsburg2014}, TinyTim \citep{Krist1995}, pysynphot \citep{Pysynphot}}

\acknowledgments We thank the anonymous referee for a constructive referee report. D.A. acknowledges support by NASA under agreement No. NNX15AD94G for the program Earths in Other Solar Systems. Support for Program number 14241 was provided by NASA through a grant from the Space Telescope Science Institute, which is operated by the Association of Universities for Research in Astronomy, Incorporated, under NASA contract NAS5-26555. Based on observations made with the NASA/ESA Hubble Space Telescope, obtained in GO program 14241 at the Space Telescope Science Institute.


\appendix
\section{Background source information}

\begin{figure}[h]
  \centering
  \includegraphics[width=0.8\textwidth]{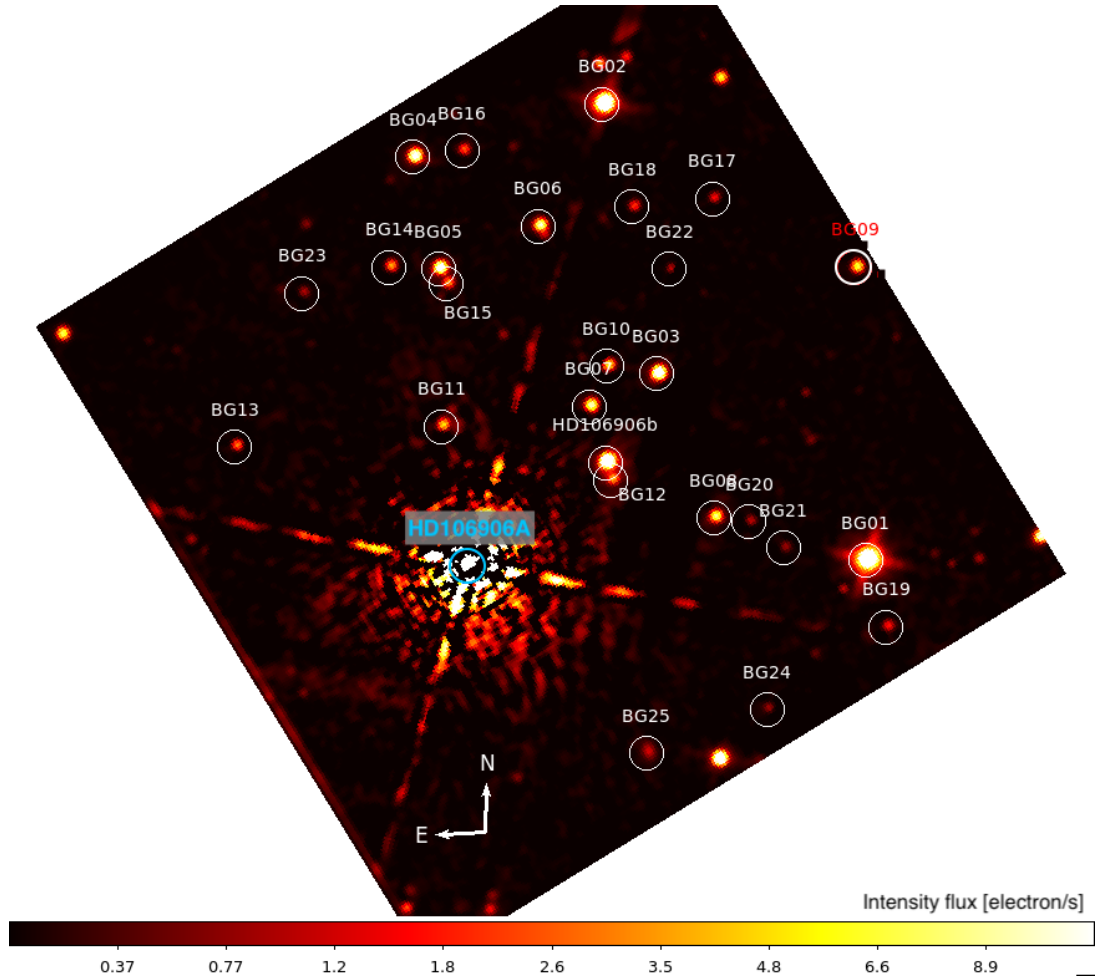}
  \caption{Illustration of sky locations of background sources in the FoV.}
  \label{fig:bck}
\end{figure}

The sky locations for background sources are illustrated in Figure \ref{fig:bck}. Table \ref{tab:bck} summarizes the information for background sources.

\begin{rotatetable*}
\begin{deluxetable*}{cccccccc}
  \footnotesize
   \tablecaption{Background Sources in the Field of View\label{tab:bck}}
  \tablehead{
    \colhead{Source ID} &
    \colhead{RA} &
    \colhead{Dec} &
    \colhead{RA} &
    \colhead{Dec} &
    \colhead{Flux$_{\mathrm{F127M}}$} &
    \colhead{Flux$_{\mathrm{F139M}}$} &
    \colhead{Flux$_{\mathrm{F153M}}$} \\
    \colhead{} &
    \colhead{2016.08} &
    \colhead{2016.08} &
    \colhead{2018.43} &
    \colhead{2018.43} &
    \colhead{\fluxunit\tablenotemark{a}} &
    \colhead{\fluxunit\tablenotemark{a}} &
    \colhead{\fluxunit\tablenotemark{a}} 
}
    \startdata
      BG01    & 12h17m51.08033s & -55d58m32.8722s & 12h17m51.08012s & -55d58m32.8703s     & 2.73e-12 & 2.44e-12 & 2.08e-12 \\
      BG02    & 12h17m52.53287s & -55d58m11.6989s & 12h17m52.53285s & -55d58m11.7027s     & 1.35e-12 & 1.20e-12 & 1.04e-12\\
      BG03    & 12h17m52.19115s & -55d58m23.9812s & 12h17m52.19254s & -55d58m23.9863s     & 2.97e-13 & 2.62e-13 & 2.23e-13\\
      BG04    & 12h17m53.48526s & -55d58m13.6424s & 12h17m53.48539s & -55d58m13.6401s     & 1.98e-13 & 1.75e-13 & 1.50e-13\\
      BG05    & 12h17m53.33002s & -55d58m18.7354s & 12h17m53.32989s & -55d58m18.7321s     & 1.77e-13 & 1.58e-13 & 1.34e-13\\
      BG06    & 12h17m52.83135s & -55d58m17.0499s & 12h17m52.83170s & -55d58m17.0504s     & 1.12e-13 & 9.89e-14 & 8.36e-14\\
      BG07    & 12h17m52.53095s & -55d58m25.2858s & 12h17m52.52897s & -55d58m25.2814s     & 1.12e-13 & 9.67e-14 & 8.65e-14\\
      BG08    & 12h17m51.86552s & -55d58m30.5719s & 12h17m51.86637s & -55d58m30.5734s     & 8.70e-14 & 7.33e-14 & 6.28e-14\\
      BG09    & 12h17m51.20143s & -55d58m19.6506s & 12h17m51.20233s & -55d58m19.6548s     & 7.27e-14 & 6.78e-14 & 6.34e-14\\
      BG10    & 12h17m52.44974s & -55d58m23.4974s & 12h17m52.44988s & -55d58m23.4991s     & 5.64e-14 & 4.96e-14 & 4.28e-14\\
      BG11    & 12h17m53.27647s & -55d58m25.8212s & 12h17m53.27557s & -55d58m25.8227s     & 5.09e-14 & 4.00e-14 & 3.28e-14\\
      BG12    & 12h17m52.40001s & -55d58m28.6510s & 12h17m52.40096s & -55d58m28.6424s     & 4.24e-14 & 3.30e-14 & 2.84e-14\\
      BG13    & 12h17m54.31592s & -55d58m26.2282s & 12h17m54.31764s & -55d58m26.2231s     & 2.95e-14 & 2.59e-14 & 2.42e-14\\
      BG14    & 12h17m53.57930s & -55d58m18.5369s & 12h17m53.57942s & -55d58m18.5387s     & 3.55e-14 & 3.29e-14 & 2.90e-14\\
      BG15    & 12h17m53.28572s & -55d58m19.4339s & 12h17m53.28540s & -55d58m19.4317s     & 3.65e-14 & 3.59e-14 & 3.10e-14\\
      BG16    & 12h17m53.23686s & -55d58m13.4621s & 12h17m53.23667s & -55d58m13.4615s     & 2.67e-14 & 2.36e-14 & 2.19e-14\\
      BG17    & 12h17m51.94942s & -55d58m16.2192s & 12h17m51.94916s & -55d58m16.2213s     & 1.71e-14 & 1.52e-14 & 1.41e-14\\
      BG18    & 12h17m52.35709s & -55d58m16.4143s & 12h17m52.35722s & -55d58m16.4142s     & 2.11e-14 & 1.85e-14 & 1.72e-14\\
      BG19    & 12h17m50.95989s & -55d58m35.9010s & 12h17m50.95988s & -55d58m35.9000s     & 1.59e-14 & 1.29e-14 & 1.35e-14\\
      BG20    & 12h17m51.68586s & -55d58m30.8703s & 12h17m51.68641s & -55d58m30.8587s     & 1.04e-14 & 1.01e-14 & 1.04e-14\\
      BG21    & 12h17m51.50254s & -55d58m32.1091s & 12h17m51.50420s & -55d58m32.1134s     & 8.87e-15 & 7.43e-15 & 6.71e-15\\
      BG22    & 12h17m52.15373s & -55d58m19.3199s & 12h17m52.15175s & -55d58m19.3031s     & 1.16e-14 & 1.03e-14 & 8.48e-15\\
      BG23    & 12h17m54.01685s & -55d58m19.5122s & 12h17m54.01515s & -55d58m19.5069s     & 9.49e-15 & 7.96e-15 & 7.61e-15\\
      BG24    & 12h17m51.54886s & -55d58m39.3202s & 12h17m51.54881s & -55d58m39.3169s     & 9.62e-15 & 7.59e-15 & 6.95e-15\\
      BG25    & 12h17m52.15398s & -55d58m40.9675s & 12h17m52.15317s & -55d58m40.9600s     & 1.23e-14 & 1.15e-14 & 1.26e-14
      \enddata
      \tablenotetext{a}{Flux is calculated by multiplying count rate and photflam. }
\end{deluxetable*}
  \end{rotatetable*}

\end{document}